\documentclass[final]{siamltex}
\usepackage{epsf}
\usepackage{amssymb}
\usepackage{amsbsy}
\usepackage{verbatim}
\usepackage[sort&compress]{natbib} 
\begin{document}
\title{A Brownian Dynamics Model of Kinesin in Three Dimensions 
Incorporating the Force-Extension Profile of the Coiled-Coil 
Cargo Tether}

\author{Paul J. Atzberger
\thanks{Rensselaer Polytechnic Institute, 
Department of Mathematics, Amos Eaton Hall, Troy, NY 12180,
Phone: 518-258-3128, Fax: 518-276-4824.  Supported by research 
grant R01 GM598775-01A1 from the National Institutes of Health.
}
\and
Charles S. Peskin
\thanks{New York University,
Department of Mathematics, New York, NY 10012;
e-mail: peskin@cims.nyu.edu.  Supported by research 
grant R01 GM598775-01A1 from the National Institutes of Health.}
}

\maketitle

\begin{abstract}
The Kinesin family of motor proteins are involved in a variety of cellular
processes that transport materials and generate force.  With recent advances
in experimental techniques, such as optical tweezers which can probe 
individual molecules, there has been an increasing interest in 
understanding the mechanisms by which motor proteins convert chemical energy 
into mechanical work.  Here we present a mathematical model for the 
chemistry and three dimensional mechanics of the Kinesin motor protein which
captures many of the force dependent features of the motor.  For the elasticity 
of the tether that attaches cargo 
to the motor we develop a method for deriving the non-linear force-extension 
relationship from optical trap data.  For the Kinesin heads, cargo, and 
microscope stage we formulate a three dimensional Brownian Dynamics model 
that takes into account excluded volume interactions.  To efficiently compute 
statistics from the model an algorithm is proposed that uses a two step protocol 
that separates the simulation of the mechanical features of the model from the 
chemical kinetics of the model.  Using this approach for a bead transported by the 
motor, the force dependent average velocity and randomness parameter 
are computed and compared with the experimental data.
\end{abstract}


\pagestyle{myheadings}
\thispagestyle{plain}
\markboth{P. ATZBERGER, C. PESKIN}
{BROWNIAN DYNAMICS MODEL OF KINESIN}

The Kinesin family of motor proteins is involved in a variety of cellular
processes including the transport of materials to the end of axons 
 ~\citep{li1999}, the controlled localization of organelles during cell 
development ~\citep{tuma1998}, and the 
separation of chromosomes during mitosis ~\citep{sharp2000},  
~\citep{karsenti2001}.
The vesicle and organelle 
transporting members of the Kinesin family interact with long microtubule
filaments of the cytoskeleton. The filaments serve the 
dual purpose of giving a cell structural support and as a highway on 
which motor proteins transport materials between distant locations 
within the cell ~\citep{thecell}.

The Kinesin motor protein consists of two homologous globular 
domains, referred to as ``heads'' ~\citep{vale1997}.  The heads 
are joined together by a long coiled-coil alpha-helix structure 
which extends to attach 
like a tether to cargo transported by the motor 
~\citep{goldstein2001}, ~\citep{kamal2002}.  The globular domains 
each have specialized regions that interact with tubulin dimers 
of the microtubules and a binding pocket for the nucleotide 
adenosine-tri-phosphate (ATP) ~\citep{woehlke1997}.

The motor moves along the microtubule track by binding and unbinding
its heads from interaction sites on the microtubule surface
~\citep{gilbert1995}.  The 
binding sites are spaced at approximately $8$-nm increments 
~\citep{svoboda1993}, ~\citep{hoenger2000}, ~\citep{amos2000c}.  
Microtubules have a polarity deriving from an asymmetry in their constituent 
monomer units ~\citep{nogales1999}, ~\citep{downing1998}.  By convention the 
end toward which almost all of the Kinesin motors move is referred to as 
the plus end of the microtubule. 

The motor is powered from energy released by breaking covalent bonds between
the second and third phosphate group of (ATP) in a process known as 
hydrolysis.  The detailed mechanism by which the 
energy released during hydrolysis is converted into mechanical work is 
presently unknown.  

Individual Kinesin motor proteins have globular domains on the order of
$10$-nm in diameter.  On this small length scale thermal fluctuations 
arising from collisions of the protein with the constituent molecules
of the surrounding solvent are significant and may play an important role in 
how the motor functions.  The fluctuations affect both the internal structure 
of each of the heads as well as their diffusion, when unbound, relative to the 
microtubule.  There have been many theoretical mechanisms proposed whereby 
the energy from hydrolysis and thermal fluctuations drive conformational 
changes, or more subtly, rectify fluctuations of protein structures to 
perform mechanical work ~\citep{howard2001}, 
~\citep{fox1998}, ~\citep{julicher1997}, ~\citep{bustamante2001}
~\citep{knight1999}, ~\citep{peskin1993}.

Crystollographic structures 
for the Kinesin molecule have been solved in a few different conformations 
when bound to the products of the ATP hydrolysis cycle or 
analogue substrates
~\citep{kozielski1997}, ~\citep{kull1996}, ~\citep{mandelkow1999}, 
~\citep{rice1999}, ~\citep{song2001}, ~\citep{kikkawa2001}, 
~\citep{sindelar2002}, ~\citep{yun2003}, ~\citep{sack1997} 
~\citep{case2000b}. 
The crystallographic structures and related mutagenesis studies yield many 
clues about which protein structures are important and how they might 
contribute to the operation of the motor protein.  

From these structures a partial picture of how the motor operates is 
emerging.  Connecting each head of Kinesin to the long coil-coiled 
structure is a sequence of approximately 15 amino acid residues referred to as 
the ``neck-linker''.  When
the head is bound to different ATP hydrolysis products this structure undergoes 
conformational changes which are thought to perform the working 
``power stroke'' of the motor which moves the lagging head closer toward 
the plus end of the microtubule ~\citep{rice1999}, ~\citep{rice2003}.  
A collection of loops and alpha 
helices have also been identified which interact with bound substrates.
These structures are thought to be important in communicating the identity 
of the substrate to distant parts of the molecule.  These ``switch'' 
structures affect such features of the motor as the binding affinity 
of a head for a microtubule and the conformation of the neck-linker
~\citep{kikkawa2001}, ~\citep{schliwa2001}, ~\citep{sablin2001}.

\begin{figure}[h*]
\centering
\epsfxsize = 5in
\epsffile[11 11 553 553]{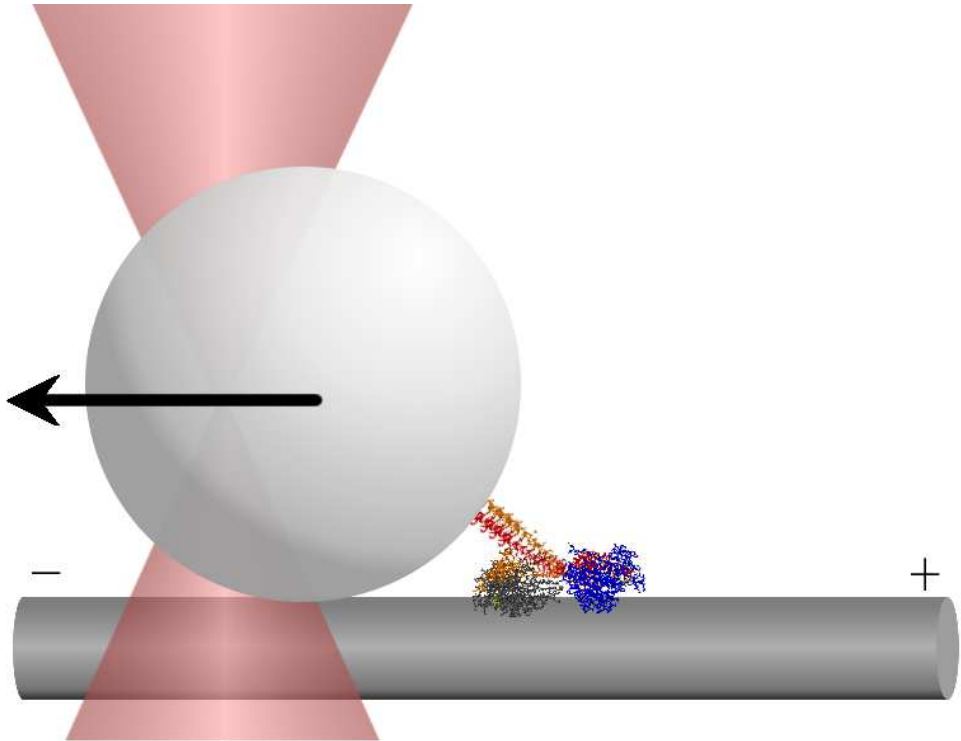}
\caption {A Schematic Representation of the Optical Trap Experiment of 
Svoboda and Block 1994.  A laser trap  for the cargo bead is generated with 
a center of focus above the microtubule.  The interaction of the laser 
light with the bead results in a restoring force that tends to 
pull the bead toward the focus of the laser.  As the motor progresses
along the microtubule toward the plus end, the bead is pulled successively 
further from the center of focus and an increasingly strong load force 
is exerted on the motor.}
\label{figure_fig_kin_bead_tow1}
\end{figure}

While crystallographic structures offer primarily a statistic geometric 
picture, optical
trap experiments have been used to study the dynamics of individual motor 
proteins as they operate.  To probe the motor, a latex bead hundreds 
of nanometers in diameter is attached to the protein through the 
long coiled-coil tether structure.  A load force is applied to the cargo bead 
and the response of its transport by the motor is observed as the load 
force is varied. See figure \ref{figure_fig_kin_bead_tow1}.  Since only 
the bead is observed, to obtain information
about the motor, a separate experiment must be performed to obtain 
the elasticity of the tether which attaches the bead to the motor 
~\citep{svobodaandblock1994}.  

Measurements from the optical trap experiments yield interesting information 
about how the motor functions.  By performing repeated experiments it is 
possible to estimate force dependent statistics of individual Kinesin 
molecules such as the response of the motor velocity to a load force.  
The experimental data also place important constraints on candidate 
mechanisms for how the motor functions, such as the number of rate limiting 
``mechanochemical'' events per 8-nm step along the microtubule.  It has even been 
possible to deduce some information about the conformational changes that 
occur.  Using high precision optical traps small systematic displacements 
of the bead have been estimated at high spatial and temporal resolution 
within the 8-nm steps of the motor
~\citep{nishiyama2001}, ~\citep{coppin1996}.

In modeling the optical trap experimental data a difficulty arises in 
modeling the motor protein.  Attempting to model at a detailed
level using methods such as molecular dynamics allows for simulation 
over only a short time relative to the time scale on which the 
motor operates and on which experimental measurements 
are made.  The overall aim of this work is to derive a 
mechanical model of the motor at a coarse scale that allows 
for efficient computation of long time-scale observables.

Many models have been proposed for Kinesin ~\citep{mogilner2001},
~\citep{peskin1995}, ~\citep{fisher2001}, ~\citep{astumian1999}, 
~\citep{maes2003}.  In these papers Kinesin is 
often described by a single mechanical degree of freedom or 
reaction coordinate in which many details concerning the geometry 
and elasticity of the motor are neglected.  
In this work we propose a three dimensional mathematical model for 
the Kinesin motor protein taking into account basic structural 
features of the motor.  Data from the optical trap experiments of 
~\citep{svobodaandblock1994}, in which the elasticity of the 
cargo tether of the motor is probed, are incorporated into the mechanics of
the model.  We show how a force-extension relationship for the tether can 
be derived from the published data.  An algorithm is then proposed that 
exploits a separation of time scales in the model to efficiently compute 
statistics that can be compared with the optical trap experimental data.

\clearpage
\pagebreak
\newpage

\clearpage
\pagebreak
\newpage

\section{Model Description}
\label{section_model_description}

Kinesin consists of two nearly homologous globular domains that form what are 
referred to as the ``heads'' of the motor protein.  Each of the heads has a 
special binding site in which ATP hydrolysis occurs and each has a separate 
structure that interacts with microtubules.  Extending from each of the heads 
is a long alpha helix that dimerizes the two heads by forming an intertwined 
coiled-coil structure.  The coiled-coil structure also acts as a tether to 
attach cargo to the motor.

\begin{figure}[h*]
\centering
\epsfxsize = 5in
\epsffile[11 11 553 553]{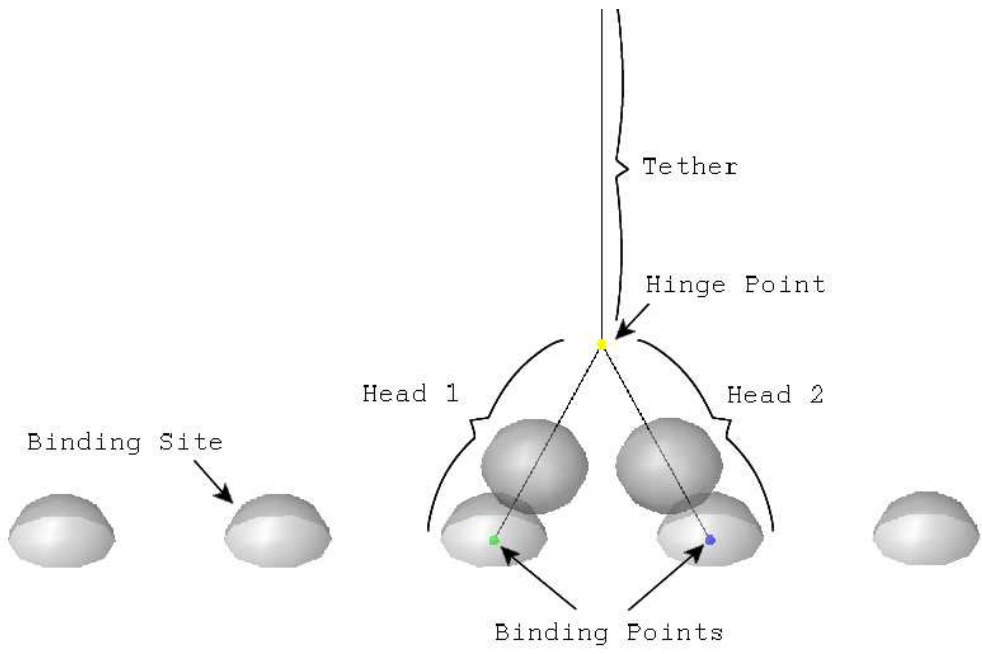}
\caption {Illustration of the Model with a Magnified View of the 
Motor}
\label{kinesin_close_view}
\end{figure}

The detailed internal geometry of the two heads of Kinesin are neglected in
the model and represented by two spherical excluded volumes.  The 
microtubule binding sites of each head are modeled by control points 
extending from the surface 
of each excluded volume.  The heads are connected to one 
another at a common point which we shall refer to as the ``hinge point''.   
The 
coiled-coil tether 
is modeled as a non-linear spring with a force-extension relationship 
derived from experimental data.  See section \ref{derive_tether} for the 
details of this procedure.  Figure \ref{kinesin_close_view} gives a 
schematic illustration of the model.

\begin{figure}[h*]
\centering
\epsfxsize = 5in
\epsffile[11 11 553 553]{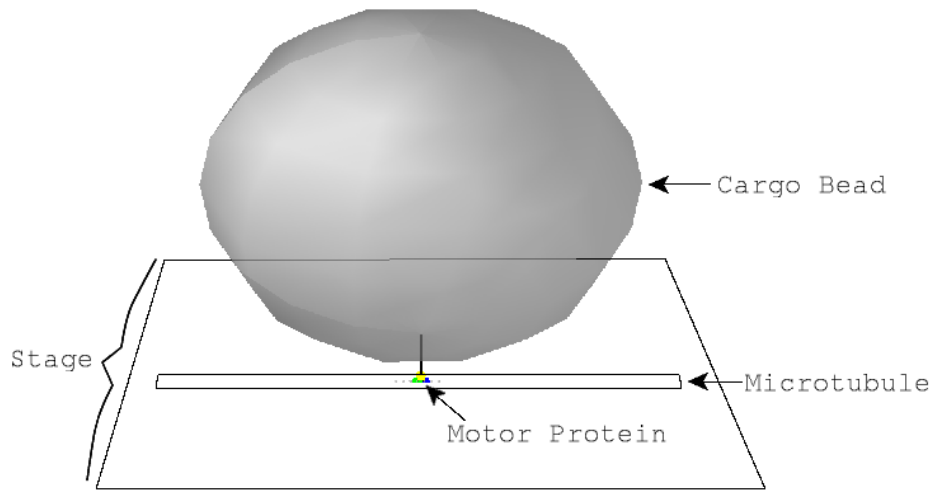}
\caption {Illustration of the Model Plotted Approximately to Scale}
\label{kinesin_far_view}
\end{figure}

A bead is connected to the motor in optical trap experiments and is modeled
by a large spherical excluded volume.  The excluded volume 
interactions of 
the bead with the microscope stage, microtubule, and the motor are accounted 
for in the model.  The interaction with the stage is enforced by a condition
that the bead not move below a given planar surface.  The detailed 
geometry of the stage mounted microtubule is neglected in the model since
on the length scale of the bead it is expected that the microtubule appears
as little more than a small ``bump'' on the stage surface only making a
minor contribution to the bead diffusion dynamics.
See figure
\ref{kinesin_far_view} for a schematic of the model where the motor, 
microtubule, and cargo bead are drawn approximately to scale.

Microtubules serve as the track on which Kinesin moves.
Typically 13 protofilaments join laterally to form a sheet that when 
rolled up forms the hollow cylindrical structure of the microtubule
~\citep{nogales1999}.  For 
dimeric Kinesins it has been found that the motor moves
along the axis of a single protofilament of the microtubule.  When the 
protofilaments are twisted so that they form a helical spiral that 
wraps around the microtubule it is found that dimeric Kinesin move in a 
similar helical spiral ~\citep{ray1993}.  This suggests that dimeric Kinesin 
binds to sites located in a regular pattern in the neighborhood of a 
protofilament.  When considering single headed Kinesin molecules a 
more complex movement along microtubules has been observed 
~\citep{berliner1995}.  

In the model we arrange the microtubule binding sites of the motor along a 
single protofilament spaced with $8$-nm increments.  We model these sites by 
hemispherical regions of radius $2$-nm that interact with the binding control
points of the Kinesin heads, see figure \ref{kinesin_close_view}.

To model the state of the motor as it progresses through the coupled 
hydrolysis and mechanical stepping cycle, we make stochastic transitions 
determined by the chemical kinetics and mechanics of the motor protein.
We postulate that each of the heads of Kinesin is in any of three 
broadly defined states.  A head can be bound to a microtubule denoted 
(B), be freely diffusing with weak affinity for the binding sites 
(W), or be freely diffusing with strong affinity for the binding sides 
(S).  The states of the motor as a whole consist of all pair combinations 
of theses affinities.  We shall discuss the details of the admissible states 
and the specific mechanochemical cycle for our model below.

In summary, the mechanical model of the motor protein consists of 
two spherical excluded volumes connected at a common hinge point.
The hinge point serves as the anchor for the tether that attaches 
cargo to the motor.  The tether is modeled by a non-linear spring.
The motor moves by successively binding and unbinding its heads 
from the track.  When a head is unbound from the microtubule there 
is a joint diffusion of the cargo bead and the heads.  When a head 
is bound to adjacent microtubule sites the the binding control 
points and the hinge point form an equilateral triangle with sides 
of length $8$-nm.  

\clearpage
\pagebreak
\newpage

\clearpage
\pagebreak
\newpage

\clearpage
\pagebreak
\newpage

\section{Reconstruction of the Tether Force-Extension Profile from 
Experimental Data}

\label{derive_tether}

Extending from the motor domains of Kinesin is a long coiled-coil
alpha helix structure which acts like a tether that attaches cargo
transported by the motor protein.  Since the forces acting on
the cargo are transmitted to the motor by the coiled-coil structure
the elasticity of the tether may play an important role how the
motor operates when transporting cargo subject to a load force.
In ~\citep{elston2000}, ~\citep{elston2000b}, ~\citep{chen2002} 
a linear force-extension relation for the tether was considered.  
It was shown that depending on the model of the motor protein 
different tether stiffnesses were optimal in the velocity attained
by the motor.  In this section we discuss a method by which a nonlinear 
force-extension relation for the coiled-coil structure can be derived 
from the data of the optical trap experiments of ~\citep{svobodaandblock1994}.

In the optical trap experiments of ~\citep{svobodaandblock1994} a bead is 
attached to an inactive Kinesin motor protein which is bound to a microtubule.  
The microtubule is mounted to a microscope stage which is moved at a fixed speed 
$V_m$.  In the experimental data only two quantities are
published.  The first quantity is the velocity ratio $r = \frac{V_b}{V_m}$
where $V_b$ is the bead velocity and $V_m$ is the stage velocity.  The second 
quantity is the component of the bead position in the direction of the 
microtubule $x_b = \mathbf{X}_b^{(1)}$, where in our notation 
$\mathbf{X}_b$ is the three dimensional position of the bead.  The experimental 
data is published as the velocity ratio $r$ plotted as a function of the bead 
position $x_b$. figure \ref{figure_tetherCompositeData}.

To obtain the force-extension profile of the tether from the one 
dimensional observations of the experiment we must determine  
the extension of the tether for each observed bead position and the 
applied force that yields this extension.  To obtain the correct relationship
we must take into account the three dimensional geometry of the experiment.

In the experiment as the stage moves the tether pulls the bead from the 
center of 
the trap which then exerts an opposing optical restoring force.  The bead 
also experiences 
random 
forces as a consequence of the thermal fluctuations of the surrounding solvent 
that cause it to diffuse in three dimensions subject to the deterministic
forces of the system.
As a consequence of the thermal fluctuations of the bead and the resolution of 
the experimental measurements, the data reflect averages of the quantities 
$r$ and $x_b$.  It can be shown that the problem of 
precisely determining 
the restoring force of the tether and the extension of the tether from 
these observations is 
a mathematically ill-posed problem.  However, if some operational 
assumptions are made an approximate force-extension profile can 
be reconstructed from the data.

\subsection{Model of the Experiment and the Reconstruction Method}
In the experiments of ~\citep{svobodaandblock1994} 
a $250$-nm latex bead was attached to the motor protein and the center of the 
optical trap was located $250$-nm above the microscope stage.  This  
arrangement has the convenient feature of keeping the bead in a 
configuration which is in contact with the surface of the microscope 
stage.  As a consequence some important simplifications can made in
the derivation.

To avoid the difficulties of modeling the thermal fluctuations of the system
we shall assume in our model that the stage velocity is slow relative to 
the time scale for the bead diffusion to reach its equilibrium distribution.
Thus at each instant we shall make the operational assumption that on average 
all forces are balanced in the system and neglect further effects of the 
thermal fluctuations.

\begin{figure}[h*]
\centering
\epsfxsize = 5in
\epsffile[109 252 485 538]{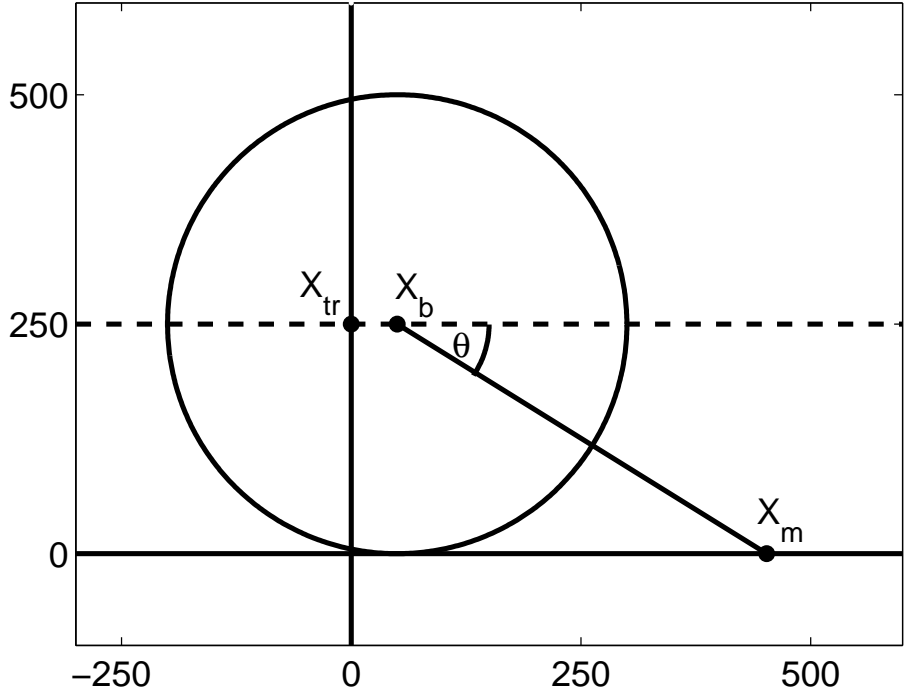}
\caption{Diagram of the  Optical Trap Experimental Setup for 
the Bead-Motor Tether.  The center of the optical trap $x_{tr}$ is 
positioned so that the bead is in contact with the microscope stage.  
The center of the bead $x_b$ in the direction of the microtubule is 
observed as the stage is moved.  The tether connecting the center of
the bead to the motor at position $x_m$ is extended under the restoring
force of the optical trap which is proportional to $x_b - x_{tr}$.  The angle 
at which the tether is connected to the bead with respect to the direction 
parallel to the microscope stage is denoted by $\theta$.  Note that the 
tether itself only consists of the segment from the motor to the surface of
the bead and has the same binding angle as that illustrated.  We also
remark that by the geometry of the setup the extension of the tether only 
depends on $x_m - x_b$.}
\label{figure_2DModelDiagram}
\end{figure}

In the experiment the motor is bound in an inactive state at position 
$\mathbf{X}_m$ to a microtubule.  The microtubule is mounted on the microscope 
stage.  A cargo bead is attached through the coiled-coil tether structure
to the motor.  The bead is subject to the restoring force toward the
center of the optical trap located at position $\mathbf{X}_{tr}$.  As a 
consequence of the geometric setup of the experiment with the bead in 
contact with the stage, when the forces are balanced, the state of the system 
is described by the one dimensional components of the positions in the 
direction of the microtubule axis.  These positions are denoted 
respectively by
$x_b = \mathbf{X}_b^{(1)}$ for the bead, 
$x_{tr} = \mathbf{X}_{tr}^{(1)}$ for the trap, and $x_m = \mathbf{X}_m^{(1)}$
for the motor.  An illustration of the geometry of the experiment is 
given in figure \ref{figure_2DModelDiagram}.

In the experiment, calibration measurements were made to obtain
the restoring force of the optical trap.  This was found to be 
approximated well by a linear spring.  In our model we shall treat the force acting 
on the bead in the direction of the microtubule arising from the optical trap 
by the linear spring $K_{tr}(x_b - x_{tr})$, in which the spring stiffness 
$K_{tr}$ has been estimated for each laser power by calibration measurements. 
The assumption of balance of forces for each configuration $x_b$ of 
the bead and 
motor $x_m$ requires that the tether force $F_{tether}$ balance the 
force of the 
optical trap in the direction of the microtubule axis.  This requires 
that the 
tether force satisfy
\begin{eqnarray}
F_{tether}\cdot \cos(\theta) = K_{tr}(x_b - x_{tr}) 
\end{eqnarray}
where $\theta$ is the angle the tether makes when binding the cargo bead
as illustrated in figure \ref{figure_2DModelDiagram}.

From this model we find that in order to obtain the tether force $F_{tether}$ 
and the extension of the tether $L$ we must determine the angle $\theta$ from
the experimental data.  This is equivalent to knowing the value of $x_m$ for 
each observation $x_b$.  We can obtain the motor position $x_m$ from the 
observations of $x_b$ and $r(x_b)$ using the following approximations.
\begin{eqnarray}
r = \frac{V_b}{V_m} \approx
\frac{\frac{\Delta{x_b}}{\Delta{t}}}
{\frac{\Delta{x_m}}{\Delta{t}}} 
\approx
\frac{\Delta{x_b}}{\Delta{x_m}}
\approx 
\frac{dx_b}{dx_m}
\end{eqnarray}

Thus we can relate $x_m$ to the observed quantities $x_b$ and $r(x_b)$ by
using the experimental data to numerically evaluate the following integral.
\begin{eqnarray}
x_m(x_b) - x_m^0
& = & 
\int_{x_b^0}^{x_b} 
\frac{\partial x_m}{\partial{x_b'}} dx_b' \\
\nonumber
& = & 
\int_{x_b^0}^{x_b} 
1/r(x_b') dx_b' 
\end{eqnarray}
We remark that this yields a well-defined function for $x_m(x_b)$ provided 
that $r(x_b) > 0$ which is indeed the case for the experimental observations.

The integration constant $x_m^0$ can be solved in terms of 
the rest-length 
$L_0$ of the tether and $x_b^0$ by
\begin{eqnarray}
x_m^0 = x_b^0 + \sqrt{(R_{bead} + L_0)^2 - R_{bead}^2}
\end{eqnarray}
We should point out that the additional term $R_{bead}$ in the term
with $L_0$ appears
because the tether is attached to the surface of the bead and not
the center of mass of the bead.

From structural considerations of the Kinesin motor protein the tether 
is estimated to have a rest length of about $L_0 = 65$-nm and the elasticity
of the tether is reported to begin to rise significantly from zero for 
$x_b^0 \approx 50$-nm in
~\citep{svobodaandblock1994}.

From the quantities $x_m$ and $x_b$ and from a trigonometric 
identity we obtain that
\begin{eqnarray}
\cos(\theta) = \frac{x_m - x_b}{L + R_{bead}}
\end{eqnarray}

We obtain from the experimentally observed quantities $x_b$ and 
the derived quantities $x_m$ the following equation for the force-extension 
profile.
\begin{eqnarray}
F_{tether} & = & \frac{L + R_{bead}}{x_m - x_b}K_{tr}(x_b - x_{tr}) \\
\nonumber
L          & = & \sqrt{(x_b - x_m)^2 + R_{bead}^2} - R_{bead} 
\end{eqnarray}

\subsection{Tether Force-Extension Reconstruction from the Experimental Data}
\label{section_tether_derivation}

In the paper of ~\citep{svobodaandblock1994} the optical trap experiments
were repeated at three laser powers $15 \mbox{mW}$, $30 \mbox{mW}$,
and $62.5 \mbox{mW}$.  For each of the experiments the restoring 
forces of the optical trap are different.  This yields information 
about the elasticity on different but potentially overlapping 
extensions of the tether.  In order to combine this information we 
derive consistency conditions that can be used to map observations 
made in one experiment to equivalent observations that would be made 
in another.  From a 
modeling point of view this allows for all of the data to be regarded 
as having been obtained in a single experiment.  Here we shall
map all of the data to an optical trap experiment formed with a 
$15 \mbox{mW}$ laser. 

There are two conditions that must be satisfied in
the experimental measurements of $r$ and $x_b$ for a 
given extension of the tether.  These requirements follow 
from the assumption that the elastic 
behavior of the tether, as it is extended, is independent of the 
stiffness of the optical trap used to probe this behavior.  The 
conditions are also a consequence of the geometry of the experiment 
in which the bead is in contact with the stage and the fact that we 
have assumed that the bead equilibrates each instant to a position 
where forces are balanced in the system.  These assumptions along
with the fact that the experiments probe identical features of the 
tether allows for data obtained from one experiment to be used to 
predict observations that would be obtained in another.  

By the geometry of the experiment 
the extension of the tether and the difference $x_b - x_m$ are in a 
one-to-one correspondence, in other words, knowledge of one determines the other.  
In fact the component of the tether force in the direction of the microtubule 
is given by a function $F_{x-tether}(x_b - x_m)$.

Since the optical trap has a linear restoring force the component of the 
optical force in the direction of the microtubule is given by 
\begin{eqnarray}
F_{tr}(x_b - x_{tr}) = -K_{tr}(x_{b} - x_{tr})
\end{eqnarray}

The condition that 
forces acting on the bead balance each instant yields the 
requirement that $F_{tr} + F_{x-tether} = 0$ which can be expressed as
\begin{eqnarray}
K_{tr}(x_b - x_{tr}) = F_{x-tether}(x_b - x_m)
\end{eqnarray}

To obtain the first consistency condition consider two experiments in 
which the tether has the same extension.  In each experiment the 
optical trap restoring force then must be the same.  This
is required so that the component of the tether force in the microtubule 
direction is balanced by the optical trap force.  Consequently, the following 
condition must be satisfied for the value $x_b^A$, measured in experiment A, 
in relation to the value of $x_b^B$, measured in experiment B.
\begin{eqnarray}
K_{tr}^A(x_b^A - x_{tr}) =  F_{x-tether}(x_b - x_m) = K_{tr}^B(x_b^B - x_{tr})
\end{eqnarray}

The second consistency condition is obtained by differentiating, with 
respect to time, the balance of force condition
which relates the optical trap restoring force to the tether 
force $F_{x-tether}(x_b - x_m)$.  We obtain after the chain-rule and 
some algebra the following relationship between the tether force in 
the direction of the microtubule and the velocity ratio 
$r = \frac{V_m}{V_b}$.

\begin{eqnarray}
\frac{r}{1 - r}K_{tr} = F'_{x-tether}(x_b - x_m)
\end{eqnarray}

We again emphasize that the force component of the tether force does not depend in
any way on the laser power used for the optical trap.  Therefore, if the tether 
has the same extension in two experiments the value $r^A$ measured in 
experiment A must satisfy the following condition with respect to the value 
$r^B$ measured in experiment B.
\begin{eqnarray}
\frac{r^A}{1-r^A}K_{tr}^A = F'_{x-tether}(x_b - x_m) = \frac{r^B}{1-r^B}K_{tr}^B
\end{eqnarray}

\begin{figure}[h*]
\centering
\epsfxsize = 5in
\epsffile[103 240 495 559]{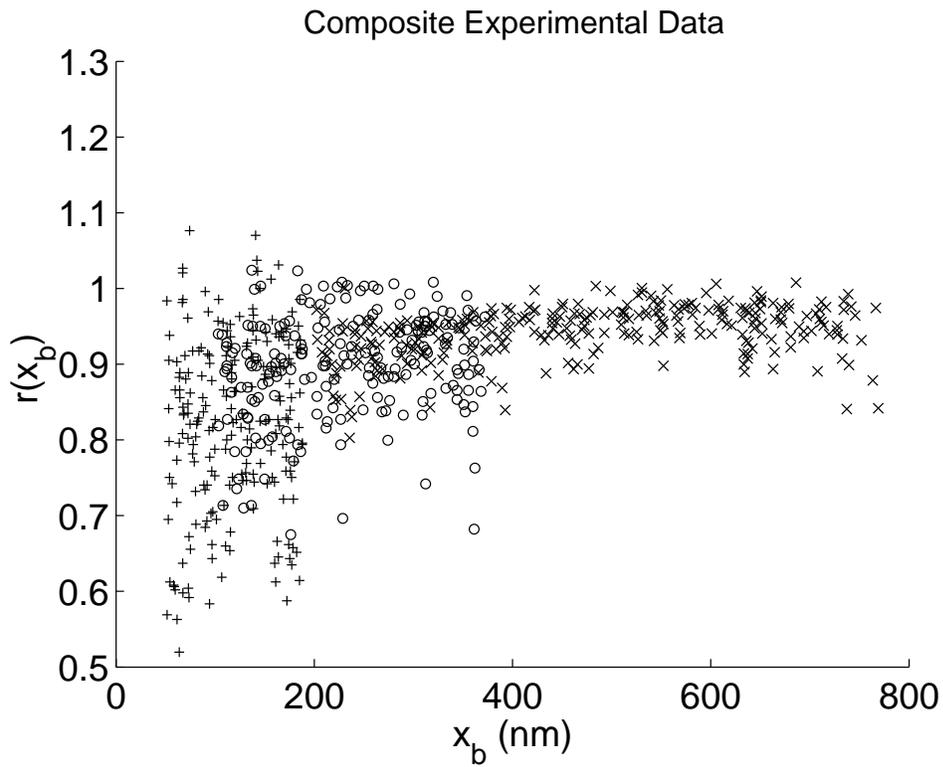}
\caption[Composite Experimental Data]
{Composite Experimental Data.  In the plot
(+) points were obtained with $15 \mbox{mW}$ laser
power, (o) with $30 \mbox{mW}$, and (x) $62 \mbox{mW}$ laser power.
The data is from ~\citep{svobodaandblock1994} and are replotted here on a 
single graph by making use of the transformation explained in the text.}
\label{figure_tetherCompositeData}
\end{figure}

\begin{figure}[h*]
\centering
\epsfxsize = 5in
\epsffile[94 232 501 558]{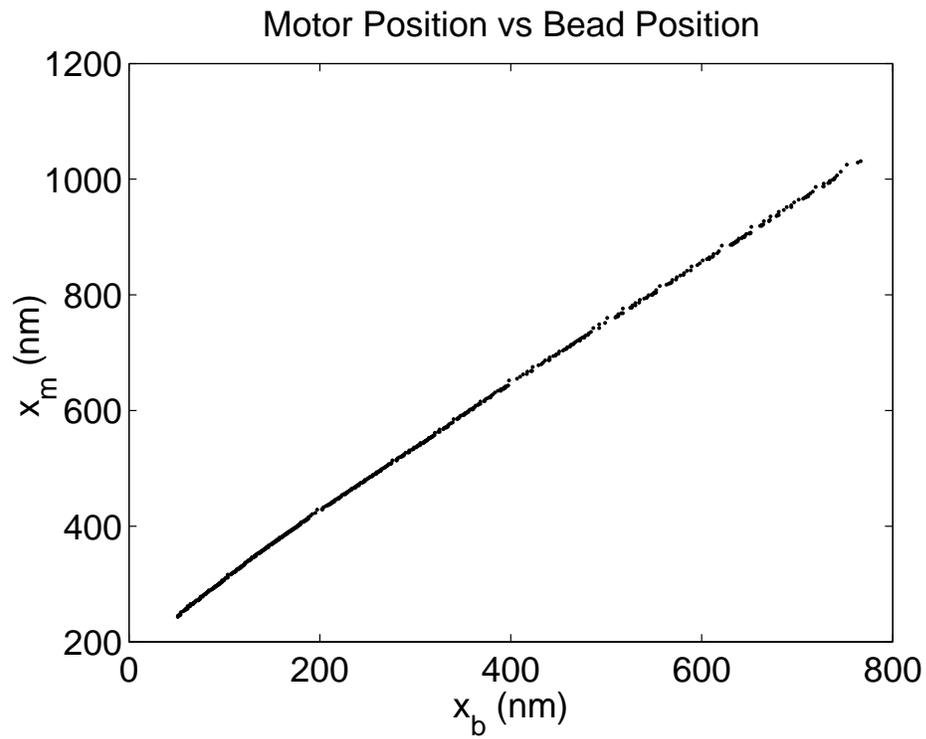}
\caption[Reconstructed Motor Position]
{Reconstructed Motor Position.  The black points show the 
reconstructed motor position $x_m$ as a function of the 
experimentally observed bead position $x_b$ obtained by 
the integration procedure described in the text.}
\label{figure_tetherX_m_vs_X_b}
\end{figure}

\begin{figure}[h*]
\centering
\epsfxsize = 5in
\epsffile[106 240 496 556]{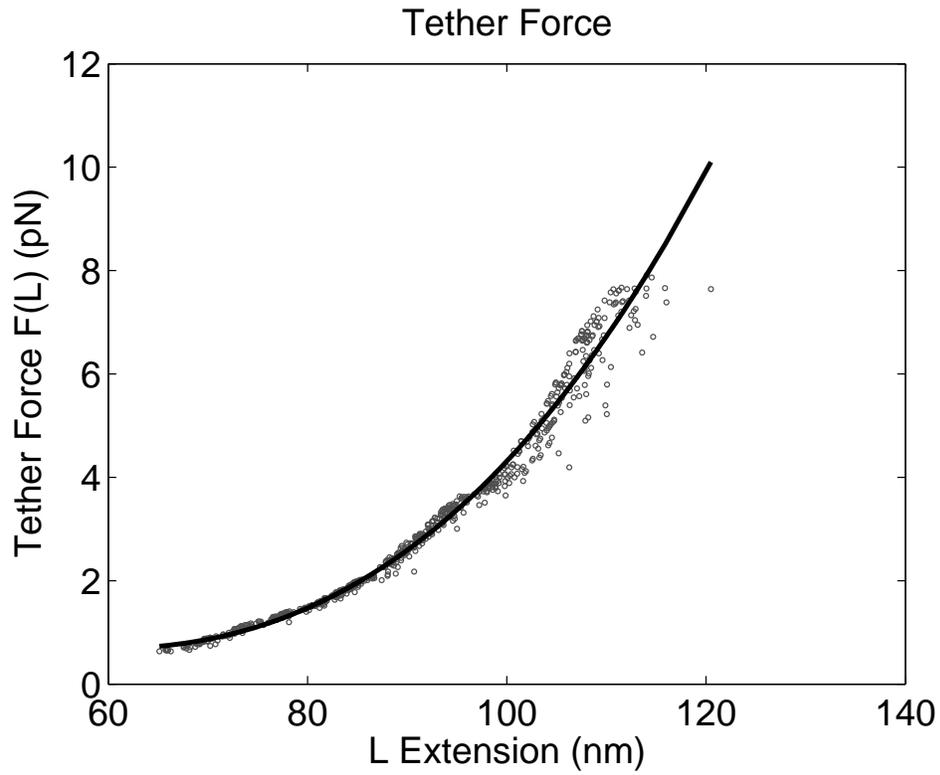}
\caption[Tether Force-Extension Profile]
{Tether Force-Extension Profile.
The plotted points are the reconstructed force-extension profile 
obtained from the procedure described in the text.  The solid curve 
is a cubic polynomial with coefficients fit by the method of 
least-squares to the reconstructed force-extension profile of the 
tether.  For small tether extensions in the model the 
force-extension profile is linearly interpolated to zero as the 
extension approaches zero.}
\label{figure_tetherForce}
\end{figure}

\begin{figure}[h*]
\centering
\epsfxsize = 5in
\epsffile[91 240 495 561]{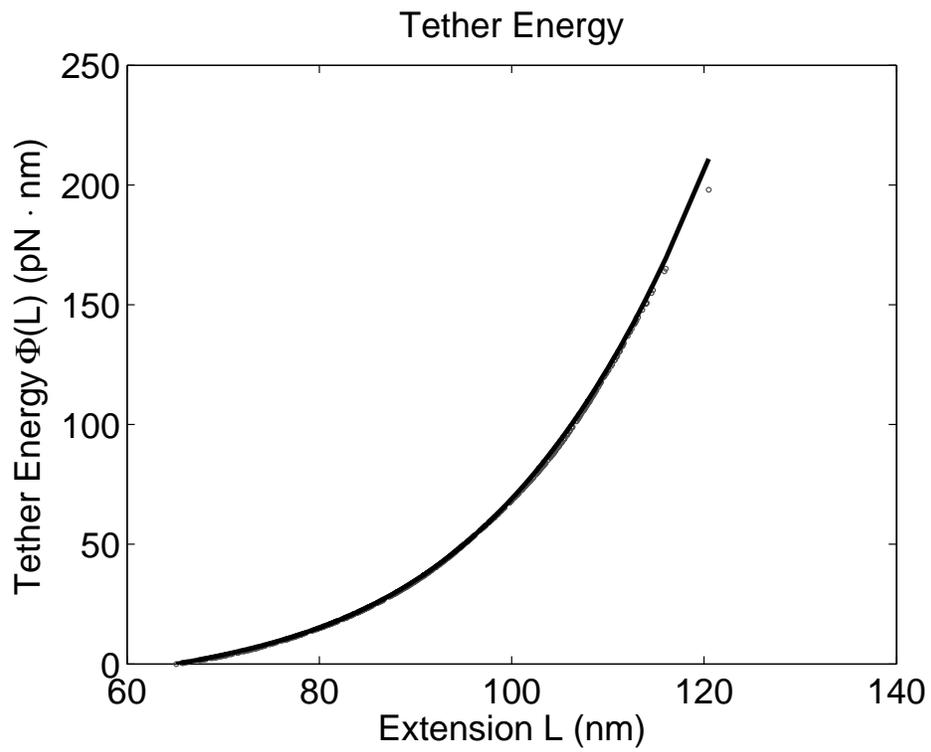}
\caption[Tether Energy-Extension Profile]
{Tether Energy-Extension Profile.
The plotted points are the reconstructed tether energy-extension profile 
obtained by numerical integration of the reconstructed tether 
force-extension profile.  The solid curve is the analytic anti-derivative 
of the cubic polynomial obtained using a least-squares fit to the tether 
force-extension profile.}
\label{figure_tetherEnergy}
\end{figure}

These two conditions allow for all of the data obtained under the three
experiments in the ~\citep{svobodaandblock1994} paper to be mapped to
equivalent observations under the experiment with an optical trap formed 
from a $15 \mbox{mW}$ laser.  The reconstruction method discussed in the 
previous section can then be applied to obtain the force-extension profile
from the composite experimental data.  The composite data of 
~\citep{svobodaandblock1994} is plotted in figure 
\ref{figure_tetherCompositeData}.  The reconstructed motor position, 
force-extension profile, and energy-extension profile are plotted 
in the figures \ref{figure_tetherX_m_vs_X_b}, 
\ref{figure_tetherForce}, \ref{figure_tetherEnergy}, respectively.

\clearpage
\pagebreak
\newpage

\clearpage
\pagebreak
\newpage

\clearpage
\pagebreak
\newpage

\clearpage
\pagebreak
\newpage

\clearpage
\pagebreak
\newpage

\clearpage
\pagebreak
\newpage

\section{The Kinetic Cycle of the Model} 

In the model we postulate that the heads of the motor protein can be in 
any of three states which are represented by their binding affinity 
for the microtubule.  The states are: bound to a microtubule (B), detached 
and diffusing with weak affinity for the microtubule (W), detached and 
diffusing with strong affinity for the microtubule (S).  The overall state
of the motor protein is then given by the binding states of each of the 
heads and the location of the bound head closest to the negative end 
of the microtubule.   For clarity in the notation we shall suppress explicit 
mention of the bound head location.  

Since we are primarily interested in the force dependent statistics of the 
optical trap experiments we shall restrict attention only to the states of 
the motor where one head is bound to the microtubule at all times.  In the
description of the model we shall not distinguish between the individual 
identity of the two heads but rather only represent their relative 
configuration.  From these considerations the states of the motor are
always in the set $\{BB,WB,BW,SB,BS\}$. 

The kinetics of the model are specified by transitions 
between these states.  For many of the states a transition 
will occur only after an exponentially distributed waiting time parameterized
by a rate constant.  The rate constant characterizes the probability per 
unit time that a transition event occurs.  If we denote the 
rate constant by $\lambda$ the waiting time is a random variable 
$\tau$ with the probability density  ~\citep{ross1995}
\begin{eqnarray}
\rho(t) = \lambda e^{-\lambda{t}}
\end{eqnarray}

In the model we do not commit to a specific correspondence between the
proposed states and the ATP hydrolysis cycle.  However, to motivate the
model we will relate our state transitions to a kinetic scheme similar
to that proposed in ~\citep{cross2004},  ~\citep{cross2000}. 

In the following description, when both heads of kinesin are bound to
the microtubule, we shall refer to the one closer to the plus end of
the microtubule as the leading head and the one closer to the minus
end of the microtubule as the trailing head.  When only one head is
bound to the microtubule, we shall not use the terminology "leading"
or "trailing" and simply refer to one as the bound head and the
other as the unbound head.

Each head of kinesin has a catalytic site which is specialized for ATP
hydrolysis.  This site may be empty, occupied by ATP, occupied by ADP
and $P_i$, or occupied by ADP only.

\begin{figure}[h*]
\centering
\epsfxsize = 5in
\epsffile[0 0 361 248]{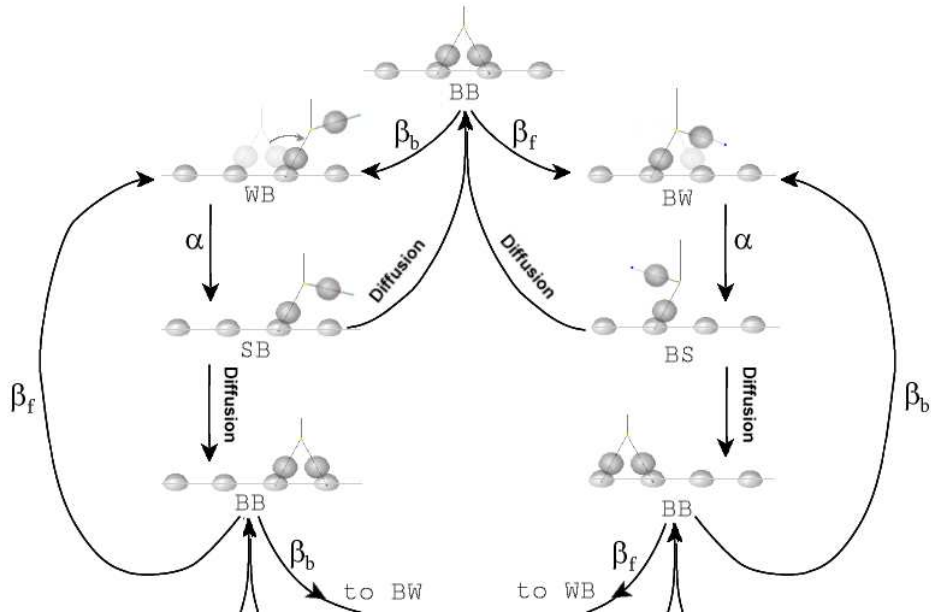}
\caption {State Transition Diagram for the Model.  Each level denotes a 
state combination for the pair of Kinesin heads.  The arrows represent
admissible transitions that can be made in the model.  Transitions that 
occur after an exponentially distributed waiting time are labeled with 
a rate constant.  Transitions that occur only after a diffusive search 
of the free head has successfully found an available binding site are 
labeled by ``diffusion''.}
\label{figure_model1_states}
\end{figure}

We begin the description of the hydrolysis cycle with one head bound
and its hydrolysis site empty, and with the other head unbound and its
hydrolysis site occupied by $ADP$.  In fact, we assume that the $ADP$
molecule is trapped in the hydrolysis site of the unbound head and
that it cannot be shed until other events occur that allow it to be
released.  This will be discussed shortly.  We refer to this state as
$WB$, meaning that one head is bound to the microtubule and the 
other head is not bound and has only weak affinity for the microtubule,
see figure \ref{figure_model1_states}.  
In particular, the unbound head will remain unbound, with $ADP$ trapped 
in its hydrolysis site, as long as kinesin remains in this state.

The next step in the cycle is the binding of $ATP$ to the bound head 
of kinesin.  Recall that the hydrolysis site of the bound head was
empty and available to bind $ATP$.  This is followed by
$ATP$ hydrolysis in the bound head.  We assume that one or the other of
these reactions, either the binding of $ATP$ or the subsequent
hydrolysis, is rate limiting and assign it the rate constant $\alpha$.
The obvious result of this step is that the bound head now has $ADP$ and
$P_i$ in its hydrolysis site, but we assume that a further result is a
change in conformation of the kinesin molecule as a whole that makes
the unbound head bind $ADP$ less strongly, and allows the unbound head
to interact strongly with the microtubule.

As a result of the above, kinesin is now in the state $SB$, in which
the unbound head will bind to the first empty kinesin-binding site on the
microtubule that it encounters.  The fast process of finding this
binding site is diffusion-limited, and therefore is not described by a
rate constant.  We obtain the statistics of this transition by direct
simulation of the stochastic dynamics of our mechanical kinesin model.

When the unbound but strongly interacting head encounters a kinesin
binding site on the microtubule it binds there.  The resulting
conformational change in the kinesin molecule as a whole causes the
shedding of $P_i$ from the other head, the one that was already bound,
which is the one on which hydrolysis has most recently occurred.

At this point we have both heads bound, so the state of kinesin is
$BB$, and both heads have $ADP$ in their hydrolysis sites.  This
situation is somewhat unstable, since the presence of $ADP$ in the
hydrolysis site weakens the interaction with the microtubule.  What
happens next is that one of the two heads sheds its ADP.  The head
that does so binds more strongly to the microtubule, while the other
head simultaneously lets go of the microtubule and binds more strongly
to its $ADP$, trapping the $ADP$ in its hydrolysis site.  The kinesin
molecule is now in the state $WB$ with which we began the description
of the hydrolysis cycle.

It is important to note, however, that there are two ways the above
coordinated reaction, release of $ADP$ from one head, and disassociation
of the other head from the microtubule, can occur, since the head that
detaches from the microtubule can be the leading head or the trailing
head.  An important assumption of our model is that the rate constants
for these two reactions are not the same.  We call $\beta_b$ the
rate constant for detachment of the trailing "back" head from the
microtubule, and $\beta_f$ the, much smaller, rate constant for
detachment of the leading "front" head from the microtubule.
Whichever head detaches, though, carries $ADP$ trapped in its hydrolysis
site, while the head that remains bound is left with its hydrolysis
site empty.

Kinesin hydrolyzes on average one ATP molecule per step along the 
microtubule 
~\citep{coy1999} and can move against load forces as great as $5 - 7pN$
~\citep{svobodaandblock1994}.  
The hydrolysis of only one ATP per step suggests there must be a mechanism 
that drives the motor forward other than the asymmetric unbinding rates 
$\beta_b$ and $\beta_f$ of the front and back heads.  If there were
no other mechanism involved we would have by mechanical symmetry that 
a head detaching from the back microtubule site would have at least a $50$\% 
probability of rebinding again to the back microtubule site.  As a consequence 
each observed mechanical step would on average require consumption of more 
than one molecule of $ATP$.

To avoid this problem we introduce into the model a mechanical 
asymmetry.  It has been reported in ~\citep{rice1999}, ~\citep{rice2003} that when 
ATP binds the motor domain a loop of about 15 amino acid residues undergoes 
a disordered to ordered transition.  This structure attaches
the motor domains to the coiled-coiled stalk and is referred 
to as the ``neck-linker''.  In the ordered state the residues of the loop 
form a partial beta sheet structure close to the motor domain.  
In the model the effect of the neck-linker is accounted 
for by a harmonic spring acting on the hinge point which introduces a 
preferred configuration in the direction of the plus end of 
the microtubule when $ATP$ is in the hydrolysis site of the bound head.
This biases the diffusion of the back head toward 
the front binding site. 

Further refinements of the model may be possible incorporating
mechanical information about the interactions of the 
neck-linker with the motor domains in ~\citep{rice1999}, 
~\citep{rice2003}.  It may also be possible to include the 
mechanism proposed by ~\citep{mogilner2001}, which involves a 
competition between a docked and undocked state
of the neck-linker, by introducing a bistable potential 
energy for the hinge point.  However, inclusion in the model 
of a potential energy with multiple minima
complicates the statistical sampling of the mechanical 
configurations.  We leave these possible refinements to 
future work.  
The kinetic cycle and mechanics of the model are summarized in figure 
\ref{figure_model1_states}.

\clearpage
\pagebreak
\newpage

\clearpage
\pagebreak
\newpage

\section{Simulation of the Model and Computation of Experimentally 
Measured Statistics}
\label{section_simulation_of_model}

From the measurements of optical trap experiments, two statistics are typically
published in the literature.  These are (i) the average velocity at which the
cargo bead is transported by the motor protein and (ii) the randomness parameter 
$q$ of the stochastic stepping process.

If we let $z_t$ be the component of the bead position along the axis of the microtubule 
at time $t$ then the average velocity of the bead is given by 
\begin{eqnarray}
v = \lim_{t \rightarrow \infty} \frac{E[z_t]}{t} 
\end{eqnarray}
where $E[\cdot]$ denotes the ensemble expectation obtained by averaging
the measurements over repeated experiments.

The randomness parameter $q$ is defined as
\begin{eqnarray}
q = \lim_{t \rightarrow \infty} \frac{\mbox{var}[z_t]}{\mbox{E}[z_t] 
\cdot{\delta}}
\end{eqnarray}
where $\mbox{var}[\cdot]$ is the ensemble variance obtained by averaging over 
repeated experimental measurements, and $\delta = 8$-nm is 
the spacing between microtubule binding sites.  

The motivation for the randomness parameter comes from a process with elementary 
chemical events having exponentially distributed waiting times with 
identical rate constants $\lambda$.  If each mechanical $8$-nm step requires 
$n$ elementary chemical events then a relatively straight-forward
calculation using the Poisson distribution for the number of elementary events 
which have occurred for the motor before the time $t$
gives $q = \frac{1}{n}$.  When the rates are not exactly the same the randomness
parameter estimates the number of slow ``rate limiting'' events
~\citep{svoboda1994b}.

In a rough sense the quantity $q$ measures the ``randomness'' of the process. 
If the number of chemical events per mechanical step is large 
$n \rightarrow \infty $ with the average time $t_0 = \frac{n}{\lambda}$ for 
a mechanical step held constant for each $n$, then the waiting time distribution 
of the mechanical step approaches the Dirac $\delta$-distribution with unit 
mass centered at the time $t_0$.  Thus as $q \rightarrow 0$ the mechanical 
stepping becomes increasingly like that of a deterministic stepping process
with delay $t_0$ between steps.

In the experiments the bead is attached through the cargo tether to an 
active Kinesin motor protein.  The position of the transported bead is 
measured with nanometer spatial accuracy and microsecond temporal 
accuracy.  At the laser powers used in the experiments which we shall consider,
the optical restoring force toward the center of focus can be approximated
to a good degree by a Hookean spring where the restoring force is a 
linear function of the displacement. 
In many of the 
experiments the laser powers are sufficiently weak that many steps of the motor 
are required before the restoring force acting on the bead changes significantly.  
The load force varies only over relatively large length scales in comparison to 
the $8$-nm step of Kinesin protein.

To estimate the load force dependence of the motor in the experiments, 
the time series of the bead position is divided into small segments which 
are determined by when the bead is within a small range of displacements from 
the optical trap center.  On each segment the statistics are computed 
by treating the load force as approximately constant.  Some of the experiments 
employ a more sophisticated
approach and use a feedback loop to move the optical trap as the motor 
progresses.  This ensures that a fixed distance is maintained
between the optical trap center and the bead and makes the load force 
to a high precision constant ~\citep{visscher1999}.

A challenge in modeling the force dependent statistics is that we must be able 
to efficiently sample the behavior of the model over many steps of the motor 
protein and over a range of load forces.  A further challenge arises from the 
range of time scales associated with the chemical kinetics and mechanics of 
the motor protein.  In general the computation of these force dependent 
statistics by directly simulating a detailed model of the experimental system 
is computationally expensive.

The approach we shall take is to simulate a coarse-grained model of Kinesin 
using a two step protocol that simulates the mechanical aspects of the 
model separately from the chemical kinetics of the model.  The geometry 
of the model was discussed in detail in section \ref{section_model_description}.
In this section we shall discuss primarily the dynamics of the model and the 
simulation method.  

In the experiment the bead and motor protein are immersed in a solvent. 
Given the small length scale of the motor protein and the cargo bead the 
Reynolds number associated with the system is small.  We shall model the 
dynamics of the model by an over-damped diffusion process where 
the $k^{th}$ mechanical degree of freedom $\mathbf{X}_t^{\{k\}}$ evolves 
according to a stochastic differential equation 
\begin{eqnarray}
d\mathbf{X}_t^{\{k\}} = -\frac{1}{\gamma_k}\nabla_{\mathbf{X}^{\{k\}}}
V(\mathbf{X}_t)dt + \sqrt{2D_k}d\mathbf{B}_t^{\{k\}}
\end{eqnarray}

The potential energy of the system is denoted by $V$ and depends 
on the configuration of the entire system, which is denoted by 
$\mathbf{X}_t$.  The details are given in the appendix.    
Associated with each degree of freedom
is an effective friction coefficient $\gamma_k$ which captures the 
dissipation of energy that occurs from the interaction of the solvent
with the bead and the heads of Kinesin.  The associated
diffusion coefficient is given by Einstein's relation, 
$D_k = \frac{K_B{T}}{\gamma_k}$, which captures the thermal 
energy that is imparted by the solvent to the bead and the heads of Kinesin
~\citep{reichl1998}.
The thermal force in the system is modeled by white noise which
we express for the $k^{th}$ mechanical degree of freedom in terms of 
increments of the standard three dimensional Brownian motion 
$\mathbf{B}_t^{\{k\}}$, which are taken as independent for each 
$k$. ~\citep{oksendal2000}.

The model is simulated numerically by using an Euler discretization of the 
stochastic differential equation ~\citep{platen1992}
\begin{eqnarray}
\mathbf{X}_k^{n + 1} = \mathbf{X}_k^{n} - \frac{1}{\gamma_k}\nabla_{\mathbf{X}_k}
V(\mathbf{X}^{n})\Delta{t} + Z_k^{n}
\end{eqnarray}

The $k^{th}$ mechanical degree of freedom at time $t_n$ is denoted by 
$X_k^{n}$.  The thermal force acting over the time increment $\Delta{t}$ 
is captured by a Gaussian random variable $Z_k$ with mean 0 and 
variance $\sqrt{2D_k\Delta{t}}$.

In the scheme the excluded volume interactions between 
the bodies are taken into account by a rejection method.  If a probabilistic 
step of the scheme results in a configuration of the system that violates the 
excluded volume constraint this step is rejected.  A new set of random 
numbers are generated for the step until an acceptable configuration is 
produced.  

From an abstract perspective, in which we view the dynamic 
trajectories as random variables drawn from the space of continuous
functions, the rejection method is equivalent to drawing trajectories 
in the subspace of admissible configurations.  The random variables are
then distributed according to the conditional probability obtained from 
the Ito measure associated with the stochastic process restricted to the 
admissible subspace.  Alternatively, from the point of view of the Ito 
diffusion process in the configuration space of the model, the rejection 
method is equivalent to imposing reflecting (no-flux) conditions on the 
boundary of the subspace of admissible configurations.

Direct simulation of the proposed model in order to generate the 
measurements used to estimate the force dependent statistics presents a 
difficulty.  The estimated rate constants of the hydrolysis cycle yield
waiting times between chemical events on the order of milliseconds.  These
waiting times are long in comparison to the time scale of the diffusive 
dynamics of the bead and the motor heads.  If the positions of the bead and 
motor heads are resolved to $0.1$ nanometer precision the time step 
must be taken on the order of $10^2$-ns.  This requires on the order of $10^4$
time steps per chemical event.  This makes the generation of even a single 
trajectory of the model spanning a $100$ chemical events computationally 
expensive requiring on the order of $10^6$ time steps.  
While it is computationally expensive to perform a direct simulation of 
the model this disparity in the time scales of the chemical kinetics and mechanics
can in fact be exploited.  

An important feature of the model is that the only diffusion limited 
event that depends directly on the mechanics of the protein is the 
rebinding of the freely diffusing head when it is in the strong 
affinity state.  While for the other chemically limited events the 
free head and the bead do diffuse, they do not influence the waiting 
for these events to occur.  In addition, since the waiting time for 
the chemically limited events in the model is long, to a good 
approximation the bead and motor have a random configuration 
distributed according to the equilibrium distribution.

\begin{figure}[h*]
\centering
\epsfxsize = 5in
\epsffile[0 0 361 248]{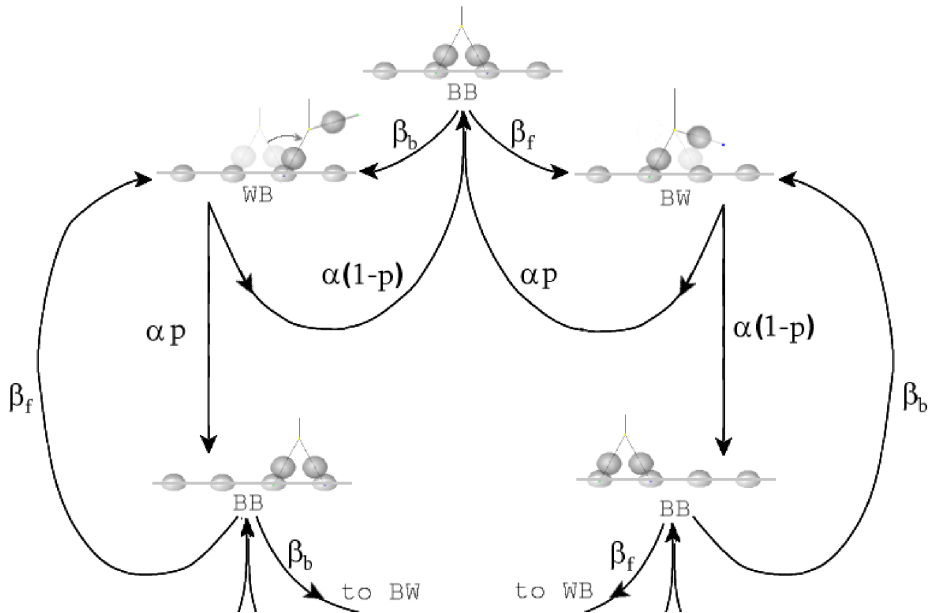}
\caption[State Transition Diagram of the Model with the Free Head Fast 
Diffusion Approximation.]
{State Transition Diagram of the Model 
with the Free Head Fast Diffusion Approximation.  Each level
depicts a state combination for the pair of Kinesin heads.  
The arrows denote the admissible transitions of the model.  Each transition 
occurs 
after an exponentially distributed waiting time with the corresponding 
rate constant denoted by the label.  The state $BS$ 
has been eliminated from the model by the fast diffusion approximation 
and the transitions
from $WB \rightarrow BS \rightarrow BB$ have been approximated by the 
transitions of the form $WB \rightarrow BB$ with rate constants 
 $\alpha p$ and
$\alpha (1 - p)$.  The factor $p = p(\mathbf{F})$ denotes for a given load
force $\mathbf{F}$ the probability that the free head binds to the site on 
the microtubule toward the plus end.}
\label{figure_model1_quasi}
\end{figure}

These insights allow for a simplification of the kinetic cycle of the model where 
the mechanical and chemical events of the motor protein can be essentially decoupled.
Since the diffusive binding of the free head occurs on a time scale much faster than 
that of the chemical events, the transition $WB \rightarrow SB \rightarrow BB$ can
be replaced by the single transition $WB \rightarrow BB$.  The transition can be
modeled by an effective rate constant $p\alpha$ for the event of the free head
binding to the front site in the direction of the plus end of the 
microtubule and $(1 - p)\alpha$ for binding to the back site.  See
figure \ref{figure_model1_quasi} for a diagram of the kinetic cycle. 

To simulate the model we use the following two step protocol:

\indent 
(i) For a given load force generate a random configuration of the bead-motor 
system in the $WB$ state distributed according to the Boltzmann equilibrium
distribution.  For each sample simulate the diffusion process of the bead
and motor until the free head binds to either the front or back site.  From 
the simulations estimate the forward binding probability. 

\begin{eqnarray}
p(\mathbf{F}) = 
\frac{\mbox{\# of times the free head binds the front site}}
{\mbox{\# of times the free head binds either the front or back site}}
\end{eqnarray}

\indent 
(ii) Simulate the simplified kinetic cycle using $p(\mathbf{F})$ from (i)
for the $WB \rightarrow BB$ transition.

This approach yields substantial computational savings.  The mechanics of
the model need only be simulated over the relatively short time required
for the free head to rebind the microtubule.  Once $p$ has been determined
the experimentally estimated statistics can be computed by simulating the 
simplified chemical kinetics of the model, which has now been reduced to 
a markov chain.  To obtain a master equation we treat the identity of each 
of the heads as indistinguishable but do make a distinction at any moment
between when a head is leading and when it is trailing.  With this convention 
the master equation is
\begin{eqnarray}
\frac{d [BB]_j}{dt} & = & p\alpha ([WB]_{j} + [BW]_{j}) 
+ (1 - p)\alpha [WB]_{j + 1} + (1 - p)\alpha [BW]_{j + 1} \\
\nonumber
& -& (\beta_b + \beta_f)[BB]_j \label{master_equ_first}\\
\nonumber
\frac{d [BW]_j}{dt} & = & \beta_f [BB]_j - \alpha [BW]_{j} \\
\nonumber
\frac{d [WB]_{j + 1}}{dt} & = & \beta_b [BB]_j - \alpha [WB]_{j + 1}  \label{master_equ_last}
\end{eqnarray}
The first symbol of the pair is the state of the back head and the second symbol
of the pair is the state of the front head.  The subscript $j$ indicates the site 
index of the bound head closest to the minus end of the microtubule.

In the implementation of the method the Metropolis algorithm is used 
to generate the random initial configuration of the bead and motor 
in the $BW$ state. 
~\citep{landau2000}.  The Forward-Euler rejection scheme proposed above
is used to simulate the over-damped diffusive dynamics with excluded volume 
interactions until the free head rebinds the microtubule.  The markov chain 
representing the 
kinetic cycle is simulated both directly by generating variates with 
exponentially distributed waiting times for each of the respective state 
transitions and by numerically solving the corresponding master's equations 
\ref{master_equ_first} - \ref{master_equ_last}.

\begin{figure}[h*]
\centering
\epsfxsize = 5in
\epsffile[65 202 524 574]{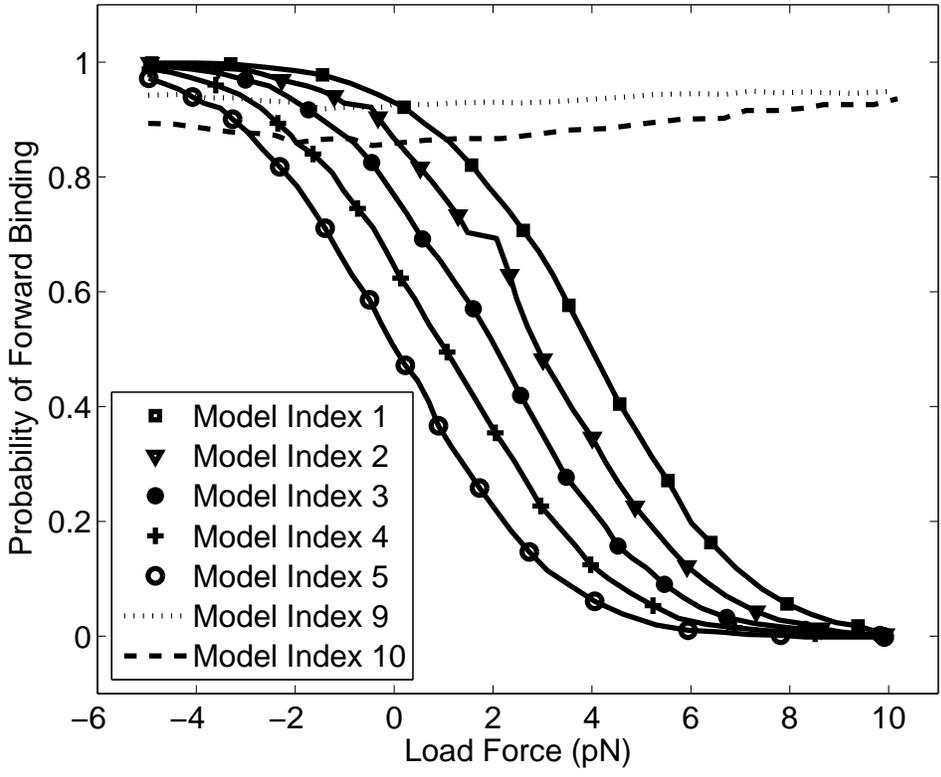}
\caption[Probability of Binding the Forward Site.]
{Probability of Binding the Forward Site.  The function $p = p(\mathbf{F})$ 
denotes for a given load force $\mathbf{F}$ the probability that the free head
binds to the site of the microtubule toward the plus end. 
The sigmoidal curves plot the probability function 
verses the load force in the direction of the minus
end of the microtubule with signed magnitude.  Thus positive load forces 
oppose the motor and negative
load forces push in the preferred direction of the motor.  The curves left to 
right plot $p(\mathbf{F})$ when the forward leaning biasing parameter of the 
model is set to $x_0^{(1)} = 0,1,2,3,4$ respectively.  The two nearly 
horizontal 
curves near the top of the figure plot the forward binding probability 
when the 
load force is taken in the direction orthogonal to the microtubule and 
parallel to
the microscope stage.}
\label{figure_kinesinSimulationVaryX0}
\end{figure}

\begin{figure}[h*]
\centering
\epsfxsize = 5in
\epsffile[51 192 556 605]{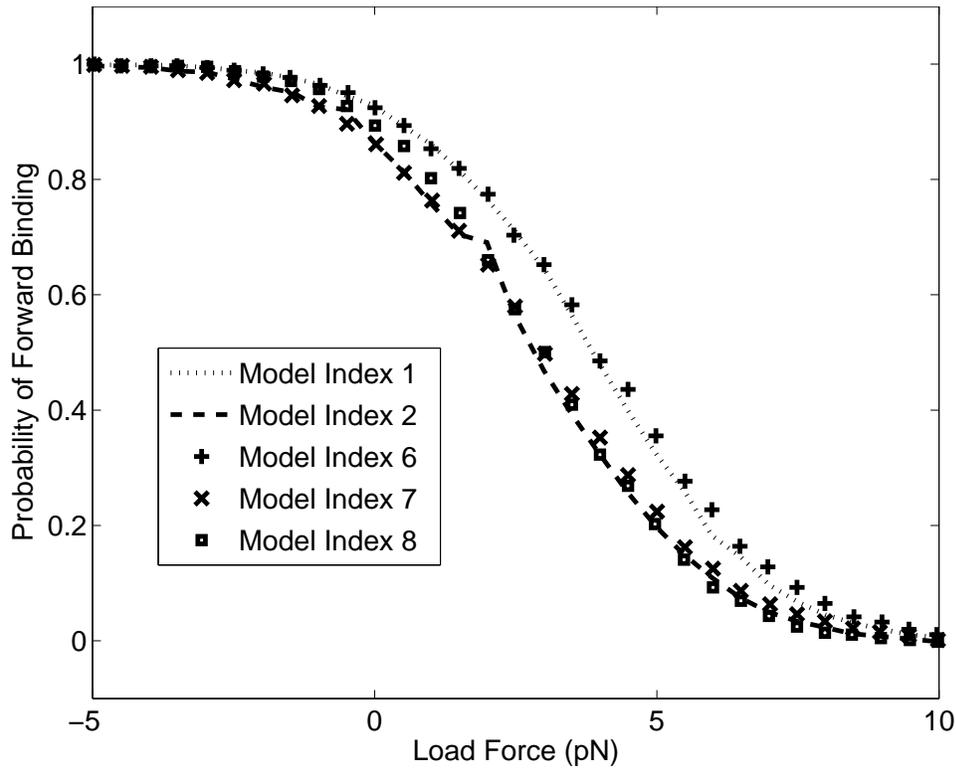}
\caption[Probability of Binding the Forward Site]
{Probability of Binding the Forward Site.
The solid sigmoidal curves left to right in the model are with forward
biasing parameter set to
$x_0^{(1)} = 3,4$ respectively.  The curves with the symbols $+$ and $\times$ 
denote the probability of binding the forward site when the nonlinear
spring that models the tether is changed to an approximating linear 
spring.  The curves with the symbols $\Box$ denote
the probability of binding the forward site when the excluded volume 
interactions are neglected in the model.  In all of the plots the load
force is taken in the direction of the minus end of the microtubule with 
signed magnitude as described in the previous figure.}
\label{figure_kinesinSimulationVaryLinearExcluded}
\end{figure}

In figures \ref{figure_kinesinSimulationVaryX0}, 
\ref{figure_kinesinSimulationVaryLinearExcluded} the front site binding probability
$p(\mathbf{F})$ is computed by varying a subset of the model parameters.  See tables
\ref{table_3Dsim_param_descr}, \ref{table_3Dsim_param_values}, 
\ref{table_3Dsim_descr}, \ref{table_3Dsim_values} for more details.  

\clearpage
\pagebreak
\newpage

\clearpage
\pagebreak
\newpage

\begin{table}[h]
\centering
\caption{The Parameters of the Three Dimensional Kinesin Model}
\vspace{0.5cm}
\label{table_3Dsim_param_descr}
\begin{tabular}{|l|l|}
\hline
Parameter & Description \\
\hline
$K_B$              & Boltzmann's constant \\
$T$                & Temperature \\
\hline
$L$                & Spacing between the binding sites. \\
$R_{bindingSite}$ & Radius of the binding sites.\\
$R_{bead}$       & Radius of the bead. \\
$K_m$            & Stiffness of the springs connecting the heads \\ 
                 & to the hinge point. \\
$R_m$            & Rest length of the spring connecting \\
                 &  the heads to the hinge. \\
$R_{glb}$        & Radius of the spherical excluded volume of \\
                 & each head. \\
$\{ a_k \}_{k = 0}^{3}$ & Coefficients of the cubic fit to the tether \\
                    & force-extension data. \\ 
$\mathbf{K}_{bias}$    & Stiffness for the restoring force of 
                      the hinge bias. \\
$\mathbf{x}_0$         & Displacement of the power stroke with the hinge point \\
                    & having the preferred position  
                      $\mathbf{X}_{bound} + \mathbf{x}_0$. \\
\hline
$\beta_b$        & Back head unbinding rate $BB \rightarrow WB$.\\
$\beta_f$        & Forward head unbinding rate $BB \rightarrow BW$.\\
$\alpha$         & Weak affinity to strong affinity transition \\
                 &  rate $WB \rightarrow SB$.\\
\hline
$\mathbf{X}_{\mbox{bead}}$  & Bead position. \\
$\mathbf{X}_{\mbox{hinge}}$ & Hinge position. \\
$\mathbf{X}_{\mbox{h1}}$    & Microtubule binding site interaction point \\ 
                         &  of head 1 position.\\
$\mathbf{X}_{\mbox{h2}}$    & Microtubule binding site interaction point \\
                         &  of head 2 position.\\
$\mathbf{F}$                & Load force acting on the bead. \\
\hline
\end{tabular}
\end{table}

\begin{table}[h]
\centering
\caption{Parameter Values for the Three Dimensional Model: The values
given here are the default values for the model when we consider variations 
of any subset of the parameters.}
\vspace{0.5cm}
\label{table_3Dsim_param_values}
\begin{tabular}{|l|l|}
\hline
Parameter & Value \\
\hline
$K_B$   & $4.142 \mbox{pN$\cdot$nm}$ \\
T       & $300 \mbox{K}$\\
\hline
$L$               & 8 \mbox{nm} \\
$R_{bindingSite}$ & 2 \mbox{nm}   \\
$R_{bead}$        & $250 \mbox{nm}$ \\
$K_m$             & $\frac{K_B{T}}{(\frac{L}{3})^2} 
                   = 0.5825 \mbox{pN/nm}$ \\
$R_m$             & $ 8 \mbox{nm}$ \\
$R_{glb}$         & $2 \mbox{nm}$ \\
$\{a_k\}_{k = 0}^{3}$ & $\begin{array}{ll} a_0 = 3.4287, & 
                          a_1 = -0.0372 \\
	                  a_2 = -0.0010, & a_3 = 1.5050 \times 10^{-5} \\
			  \end{array}$ \\
$\mathbf{K}_{bias}^{(1)}$   & $1 \mbox{pN/nm}$ \\
$\mathbf{K}_{bias}^{(2)}$   & $1 \mbox{pN/nm}$ \\
$\mathbf{K}_{bias}^{(3)}$   & $0 \mbox{pN/nm}$ \\ 
$\mathbf{x}_0^{(1)}$   & $4 \mbox{nm}$\\ 
$\mathbf{x}_0^{(2)}$   & $0 \mbox{nm}$\\ 
$\mathbf{x}_0^{(3)}$   & $0 \mbox{nm}$\\ 
\hline
$\beta_b$        & $102.5 \mbox{s}^{-1}$ \\
$\beta_f$        & $2.5 \mbox{s}^{-1}$   \\
$\alpha$         & $400 \mbox{s}^{-1}$   \\
\hline
\end{tabular}
\end{table}

\clearpage

\begin{table}[h]
\centering
\caption{Description of the Simulations of the Three Dimensional Model}
\vspace{0.5cm}
\label{table_3Dsim_descr}
\begin{tabular}{|l|l|}
\hline
Model Index & Description \\
\hline
1  & Forward bias of $x_0^{(1)} = 4$.\\          
2  & Parameters chosen to match one dimensional \\ 
   & model when possible with $x_0^{(1)} = 3$. \\                
3  & Forward bias of $x_0^{(1)} = 2$. \\            
4  & Forward bias of $x_0^{(1)} = 1$. \\            
5  & No forward bias $x_0^{(1)} = 0$. \\           
6  & Linear spring used instead of cubic model  \\  
   & of the tether with $x_0^{(1)} = 4$. \\                        
7  & Linear spring used instead of cubic model \\ 
   & of the tether with $x_0^{(1)} = 3$.\\                       
8  & Lacks excluded volume and has forward bias of $x_0^{(1)} = 3$.\\ 
9  & Sideways force applied to model with $x_0^{(1)} = 4$. \\       
10 & Sideways force applied to model with $x_0^{(1)} = 3$. \\ 

\hline
\end{tabular}
\end{table}

\begin{table}[h]
\centering
\caption{Parameter Values used for each Simulation of the 
Three Dimensional Model}
\vspace{0.5cm}
\label{table_3Dsim_values}
\begin{tabular}{|l|l|l|l|l|}
\hline
Model Index & $x_0^{(1)}$ & Excluded Volumes & Tether Type & Force Direction \\
\hline
1   & 4 & yes & cubic  & opposed  \\ 
2   & 3 & yes & cubic  & opposed  \\ 
3   & 2 & yes & cubic  & opposed  \\ 
4   & 1 & yes & cubic  & opposed  \\ 
5   & 0 & yes & cubic  & opposed  \\ 
6   & 4 & yes & linear     & opposed  \\ 
7   & 3 & yes & linear     & opposed  \\ 
8   & 3 & no  & cubic  & opposed  \\ 
9   & 4 & yes & cubic  & sideways \\ 
10  & 3 & yes & cubic  & sideways \\ 
\hline
\end{tabular}
\end{table}

\clearpage
\pagebreak
\newpage

\clearpage
\pagebreak
\newpage

\clearpage
\pagebreak
\newpage

\section{Comparison with Optical Trap Data}
In this section we compare the simulation results with the experimental data 
of ~\citep{svobodaandblock1994}, 
~\citep{block2003}, 
~\citep{coppin1997}, 
and 
~\citep{visscher1999}.  We also compare the three
dimensional model proposed here with its one-dimensional linear-spring  
counterpart proposed in ~\citep{peskin1995}.  

\begin{figure}[h*]
\centering
\epsfxsize = 5in
\epsffile[111 238 489 539]{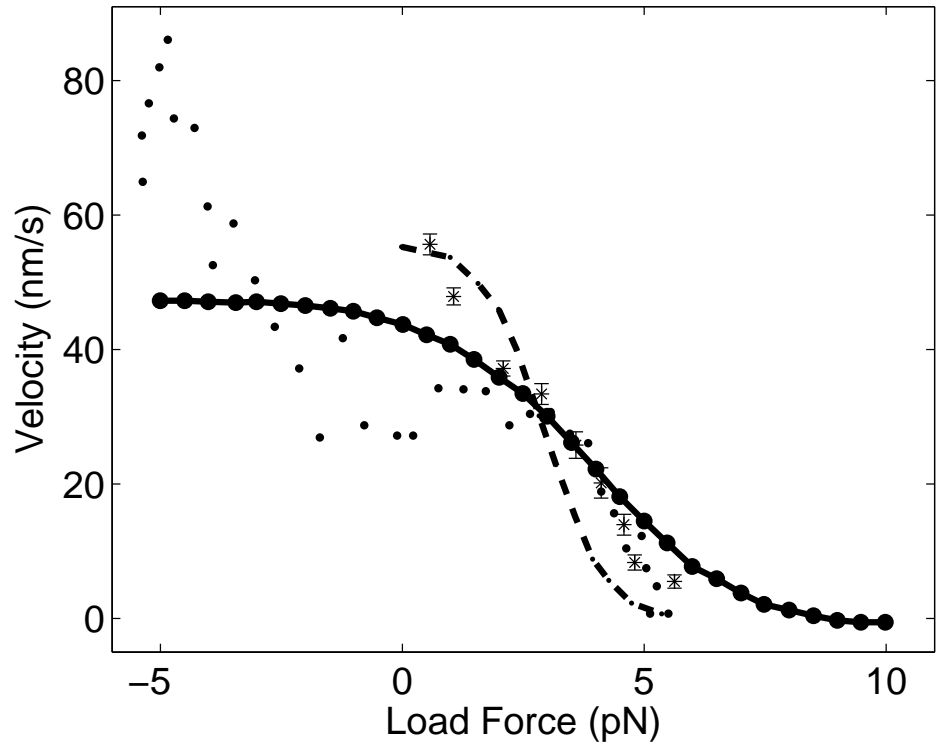}
\caption[Comparison of the Experimental Force-Velocity Data of
~\citep{coppin1997} and ~\citep{visscher1999} at $5 \mu M$ with the Model]
{Comparison of the Experimental Force-Velocity Data of 
~\citep{coppin1997} and ~\citep{visscher1999} at $5 \mu M$ with the 
Model.  The data points with the $*$ symbol and error bars are replotted from 
the paper of ~\citep{visscher1999}.  The data points with the $\bullet$ symbol
are replotted from the paper of ~\citep{coppin1997}.
The dashed curve plots the force-velocity profile of the one 
dimensional Kinesin model proposed in ~\citep{peskin1995}.  The 
solid curve plots the force-velocity profile of the best fit of the 
three dimensional model proposed in this paper.}
\label{figure_plotExpFVCompare3}
\end{figure}

Before making these comparisons we should remark that the experimental data 
available for the motor protein Kinesin is of a somewhat limited nature. 
For example significant differences appear in the literature for published
data of experiments under similar conditions.
In ~\citep{coppin1997} and ~\citep{visscher1999} the force-velocity
statistics are computed for Kinesin at a $5 \mu\mbox{M}$ ATP 
concentration.  As illustrated in the figure \ref{figure_plotExpFVCompare3}
the experimental data in the two experiments differ significantly 
especially in the range of negative load forces that push the motor 
in the direction of the plus end of the microtubule.  

Some aspects of the 
experiment that may account for the discrepancies include the use of different 
optical trap techniques which may probe the features of the motor 
through the cargo bead differently and the use of different 
biological sources to obtain the Kinesin motor proteins.  In 
~\citep{visscher1999} the optical trap center is moved as the motor progresses
to maintain an approximately fixed distance from the bead to obtain a 
constant load force to high precision.  In ~\citep{coppin1997} the optical trap 
is formed at a fixed position.  

As a result of these limitations the comparison 
we make with the experimental data is largely qualitative and primarily shows
some of the changes in the motor statistics that can arise from the three
dimensional mechanics of the motor.  Our aim in comparing with the data
is to show that the general features of the model are reasonable.

\clearpage
\pagebreak
\newpage

\begin{figure}[h*]
\centering
\epsfxsize = 5in
\epsffile[95 236 489 539]{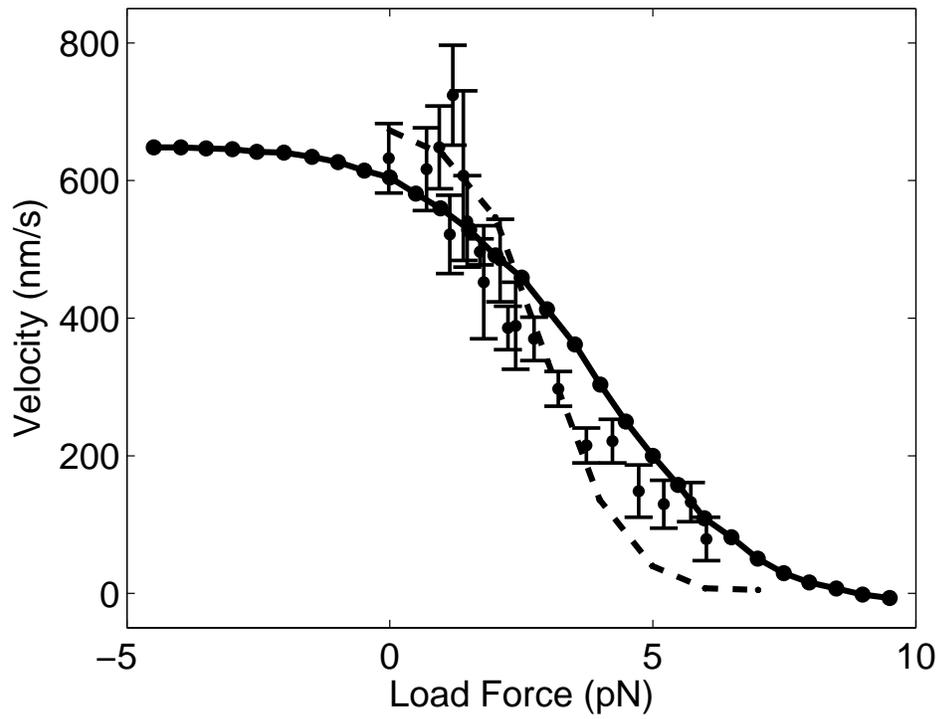}
\caption[Comparison of the Experimental Force-Velocity Data of 
~\citep{svobodaandblock1994} at 1 mM with the Model]
{Comparison of the Experimental Force-Velocity Data of 
~\citep{svobodaandblock1994} at 1 mM with the Model.  
The data points
with error bars are replotted from the paper of 
 ~\citep{svobodaandblock1994}. 
 The dashed curve plots the 
force-velocity profile of the one dimensional Kinesin model 
proposed in ~\citep{peskin1995}.  The solid curve plots the 
force-velocity profile of the best fit of the three dimensional 
model proposed in this paper.}
\label{figure_plotExpFVCompare1}
\end{figure}

\clearpage
\pagebreak
\newpage

\clearpage
\pagebreak
\newpage

\begin{figure}[h*]
\centering
\epsfxsize = 5in
\epsffile[90 236 484 540]{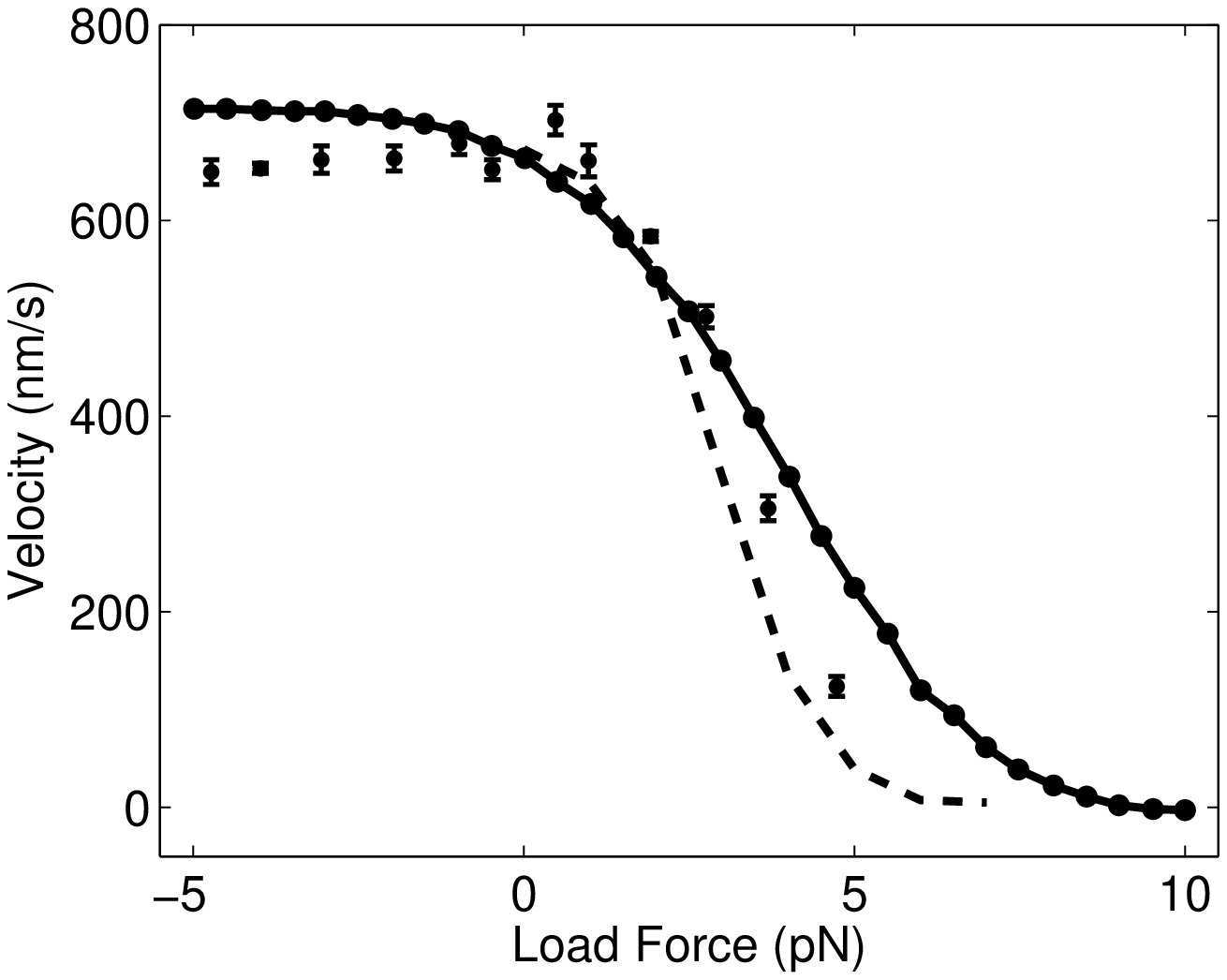}
\caption[Comparison of the Experimental Force-Velocity Data of 
~\citep{block2003} at 1.6 mM with the Model]
{Comparison of the Experimental Force-Velocity Data of 
~\citep{block2003} at 1.6 mM with the Model. 
The data points
with error bars are replotted from the paper of 
~\citep{block2003}.
The dashed curve plots the 
force-velocity profile of the one dimensional Kinesin model 
proposed in ~\citep{peskin1995}.  
The solid curve plots the 
force-velocity profile of the best fit of the three dimensional 
model proposed in this paper.}
\label{figure_plotExpFVCompare2}
\end{figure}

\clearpage
\pagebreak
\newpage

\clearpage
\pagebreak
\newpage

\clearpage
\pagebreak
\newpage

All of the experimental data was fit using the $p(\mathbf{F})$ computed
in the previous section for the three dimensional model with index 1.
Overall we find the model fit the data only moderately well.  The data set 
with 
the most discrepancy was the force velocity statistics of 
~\citep{coppin1997} in which the negative load force pushes the motor toward 
the plus end of the microtubule.  As mentioned in the beginning of this
section these data differ from ~\citep{visscher1999}.  Also in the 
experiment of 
~\citep{block2003} the trend of an increasing velocity for the motor as the 
forward load force increases was not found.  This suggests that there may be 
some features of the experimental techniques that can account for the 
difference, although it can not be ruled out that this may in fact be an 
intrinsic feature of the Kinesin motor proteins used in the experiment.  For 
the data set of ~\citep{coppin1997} we fit the model to the experimental data 
sets in the range where they tended to agree with ~\citep{visscher1999}.  

We find that much of the experimental data for the force velocity statistic
appears to fall in a range between the three dimensional model 
and one dimensional model.  As discussed in the beginning of this section, 
the force velocity statistic of the model is proportional up to a 
translation to the function $p(\mathbf{F})$.  The probability $p(\mathbf{F})$ of 
binding the forward site in the three dimensional model makes a more gradual 
transition from $1$ to $0$ as the load force increases than in the one 
dimensional model.  This indicates that the probability of the free head 
binding a forward site is much more sensitive to the load force in the one 
dimensional model.  One approach to reduce the discrepancy with the 
experimental data would be to construct a mechanical model of the motor 
protein for which the sensitivity of $p(\mathbf{F})$ to the load force falls 
between that of the proposed one dimensional and three 
dimensional models.  

An alternative hypothesis to account for the discrepancy with the 
experimental data is that the parameters associated 
with the chemically regulated steps of the motor depend on the load force.  
Both of these aspects of the model can be adjusted to fit the force velocity 
statistic but additional experimental data, such as the measurement of 
the randomness parameter, places important constraints on how this may be 
done.  One approach to distinguishing between these two alternatives is to 
require that the model fit the force variance or randomness parameter of
the experiments.  

In ~\citep{visscher1999} and ~\citep{block2003} the randomness parameter was
computed experimentally.  Below we show for the three dimensional model a
fit performed simultaneously for the randomness parameter and the force 
velocity statistic of ~\citep{block2003}.  The fit to the force 
velocity statistic was shown in the previous figures.  

\clearpage
\pagebreak
\newpage

\begin{figure}[h*]
\centering
\epsfxsize = 5in
\epsffile[98 236 483 536]{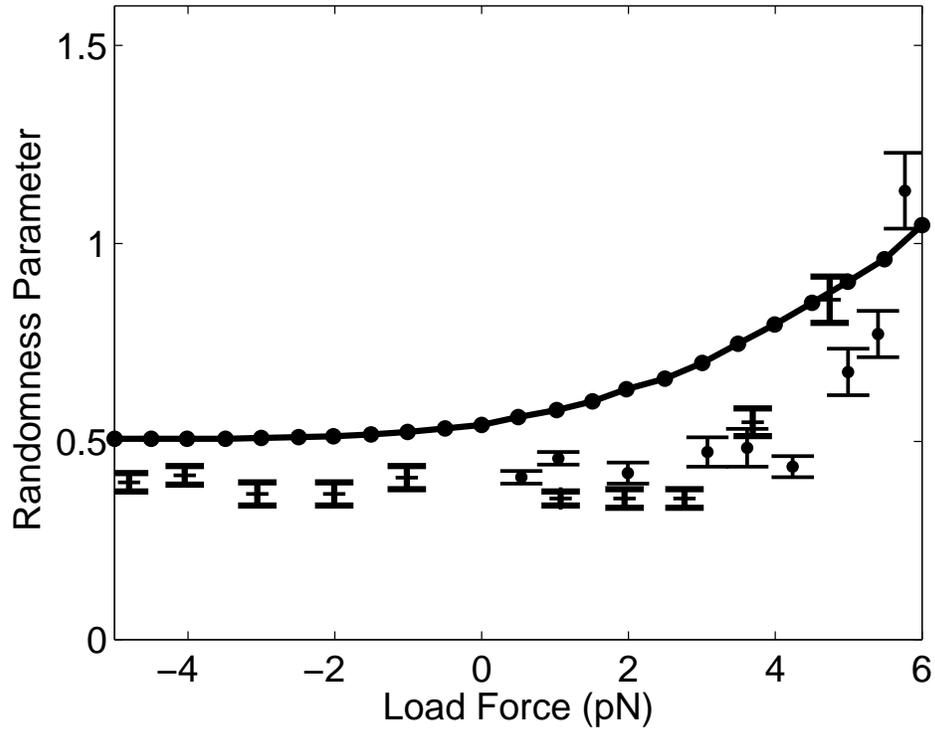}
\caption[Comparison of the Experimental Data for the Randomness
Parameter of ~\citep{block2003} at 1.6 mM and ~\citep{visscher1999} at 2 mM
with the Model]
{Comparison of the Experimental Data for the Randomness
Parameter of ~\citep{block2003} at 1.6 mM and ~\citep{visscher1999} at 2 mM
with the Model.  The data points with the $+$ symbol and 
error bars are replotted from ~\citep{block2003}.  The data points with 
the $\bullet$ symbol and error bars are replotted from ~\citep{visscher1999}.
The solid curve plots the randomness profile of the three dimensional model 
proposed in this paper fit to the force-velocity data of ~\citep{block2003}.}
\label{figure_plotExpRandCompare2}
\end{figure}

\clearpage
\pagebreak
\newpage

We find that the three dimensional model tends to have a higher 
randomness parameter but follows the same general trend. 
In the model there is a limit to how small the randomness parameter
can become when the parameters of the model are nontrivial. 
The randomness parameter can be shown to have a minimum of $\frac{1}{2}$ 
when the probability of taking a forward step is one $p = 1$.
For most of the load forces the  randomness parameter of the model is 
about $0.5$ and that of the experimental data is $0.4$.  This suggests that 
there may be  additional states of Kinesin not accounted for in the model 
that have collectively a non-negligible but moderately fast 
transition rate on the time scale of the chemically regulated steps of the 
model.  For example if we include in the model the shedding of the 
hydrolysis product $P_i$ after $ATP$ binds one of the bound Kinesin 
heads it would be expected that this does not change many of the trends 
observed in the statistics of the model but this additional exponentially 
distributed waiting time would result in an overall lowering of the 
randomness parameter.

For the sideways load forces the predictions of the model do not 
agree with the trends observed experimentally.  In the 
experimental observations of ~\citep{block2003} the velocity of 
the motor decreases as the sideways load increases.  Interestingly,
while contrary to the experimental observations of Kinesin the three
dimensional mechanical model has a velocity that is found to increase 
slightly with sideways load forces.

\section{Conclusion}
In this paper we have derived from optical trap experimental data the 
force-extension profile of the coiled-coil tether than attaches cargo to 
the Kinesin motor protein.  This was accomplished by formulating a theoretical
model of the experiment that allows the extension and the restoring force 
of the tether to be deduced from the observations of the cargo bead 
position and the velocity ratio of the bead relative to the moving stage.
We then proposed a three dimensional mechanical 
model of Kinesin incorporating the reconstructed force-extension profile 
into the model.  Force dependent statistics were computed for the model
by a procedure that exploits a separation of times scales between the 
diffusion time scales of the cargo bead and Kinesin heads and the 
time scales of the chemical kinetics of the motor.  The statistics obtained 
were then compared with experimental data.

\section{Acknowledgments}
Both authors were supported by research grant R01 GM59875-01A1 from the 
National Institutes of Health.  We are indebted to George Oster for introducing 
us to the problem of modeling Kinesin, and for helpful discussions in 
connection with this work.

\bibliography{finalManuscript_arXiv_org}

\begin{thebibliography}{10}

\bibitem{thecell}
{\sc Bruce Alberts, Alexander Johnson, Julian Lewis, Martin Raff, Keith
  Roberts, and Peter Walker}, {\em Molecular Biology of the Cell}, Garland
  Publishing, 2002.

\bibitem{amos2000c}
{\sc L.~A. Amos}, {\em Focusing-in on microtubule}, Current Opinion in
  Structural Biology, 10 (2000), pp.~236--241.

\bibitem{astumian1999}
{\sc R.~D. Astumian and I.~Derenyi}, {\em A chemically reversible brownian
  motor: Application to kinesin and ncd}, Biophysical Journal, 77 (1999),
  pp.~993--1002.

\bibitem{berliner1995}
{\sc E.~Berliner}, {\em Failure of a single-headed kinesin to track parallel to
  microtubule protofilaments}, Nature, 373 (1995), pp.~718--721.

\bibitem{block2003}
{\sc S.M. Block, C.L. Asbury, J.W. Shaevitz, and M.J. Lang}, {\em Probing the
  kinesin reaction cycle with a 2d optical force clamp}, Proc. Natl. Acad. Sci.
  USA, 100 (2003), pp.~2351--2356.

\bibitem{bustamante2001}
{\sc C.~Bustamante, D.~Keller, and G.~Oster}, {\em The physics of molecular
  motors}, Acc. Chem. Res., 34 (2001), pp.~412--420.

\bibitem{case2000b}
{\sc R.~B. Case, S.~Rice, C.~L. Hart, B.~Ly, and R.D. Vale}, {\em Role of the
  kinesin neck linker and catalytic core in microtubule-based motility},
  Current Biology, 10 (2000), pp.~157--160.

\bibitem{chen2002}
{\sc Y.~Chen, B.~Yan, and R.~J. Rubin}, {\em Fluctuations and randomness of
  movement of bead powered by a single kinesin molecule in a force-clamped
  motility array: Monte-carlo simulations}, Biophysical Journal, 83 (2002),
  pp.~2360--2369.

\bibitem{coppin1996}
{\sc C.M. Coppin, J.~T. Finer, J.~A. Spudich, and R.~D. Vale}, {\em Detection
  of sub-8-nm movements of kinesin by high-resolution optica-trap microscopy},
  Proc. Natl. Acad. Sci. USA, 93 (1996), pp.~1913--1817.

\bibitem{coppin1997}
{\sc C.M. Coppin, D.W. Pierce, L.~Hsu, and R.D. Vale}, {\em The load dependence
  of kinesin's mechanical cycle}, Proc. Natl. Acad. Sci. USA, 94 (1997),
  pp.~8539--8544.

\bibitem{coy1999}
{\sc D.~L. Coy, M.~Wagenbach, and J.~Howard}, {\em Kinesin takes one 8-nm step
  for each atp that it hydrolyzes}, J. Biol. Chem., 274 (1999), pp.~3667--3671.

\bibitem{cross2004}
{\sc R.~A. Cross}, {\em The kinetic mechanism of kinesin}, TRENDS in
  Biochemical Sciences, 29 (2004), pp.~301--307.

\bibitem{cross2000}
{\sc R.~A. Cross, I.~Crevel, N.~J. Carter, M.~C. Alonso, K.~Hirose, and L.~A.
  Amos}, {\em The conformational cycle of kinesin}, Phil. Trans. R. Soc. Lond.
  B, 355 (2000), pp.~459--464.

\bibitem{downing1998}
{\sc K.~Downing and E.~Nogales}, {\em Tubulin and microtubule structure},
  Current Opinion in Cell Biology, 10 (1998), pp.~16--22.

\bibitem{elston2000}
{\sc T.~C. Elston and C.~S. Peskin}, {\em The role of protein flexibility in
  molecular motor function: coupling diffusion in a tilted periodic potential},
  SIAM Journal of Applied Mathematics, 60 (2000), pp.~842--867.

\bibitem{elston2000b}
{\sc T.~C. Elston, D.~You, and C.~S. Peskin}, {\em Protein flexibility and
  correlation ratchet}, SIAM Journal of Applied Mathematics, 61 (2000),
  pp.~776--791.

\bibitem{fisher2001}
{\sc M.~E. Fisher and A.~B. Kolomeisky}, {\em Simple mechanochemistry describes
  the dynamics of kinesin molecules}, PNAS, 98 (2001), pp.~7748--7753.

\bibitem{fox1998}
{\sc R.~F. Fox}, {\em Rectified brownian movement in molecular and cell
  biology}, Physical Review E, 57 (1998), pp.~2177--2203.

\bibitem{gilbert1995}
{\sc S.~Gilbert and K.~Johnson}, {\em Pathway of processive atp hydrolysis by
  kinesin}, Nature, 373 (1995), pp.~671--676.

\bibitem{goldstein2001}
{\sc L.~S.~B. Goldstein}, {\em Molecular motors: from one motor many tails to
  one motor many tales}, TRENDS in Cell Biology, 11 (2001), pp.~477--482.

\bibitem{hoenger2000}
{\sc A.~Hoenger, M.~Thormahlen, R.~Diaz-Avalos, M.~Doerhoefer, K.~N. Goldie,
  J.~Muller, and E.~Mandelkow}, {\em A new look at the microtubule binding
  patterns of dimeric kinesins}, Journal of Molecular Biology, 297 (2000),
  pp.~1087--1103.

\bibitem{howard2001}
{\sc J.~Howard}, {\em Mechanics of Motor Proteins and the Cytoskeleton},
  Sinauer Associates, 2001.

\bibitem{julicher1997}
{\sc F.~Julicher, A.~Ajdari, and J.~Prost}, {\em Modeling molecular motors},
  Reviews of Modern Physics, 69 (1997), pp.~1269--1281.

\bibitem{kamal2002}
{\sc A.~Kamal and L~SB Goldstein}, {\em Principles of cargo attachments to
  cytoplasmic motor proteins.}, Current Opinion Cell Biology, 14 (2002),
  pp.~63--68.

\bibitem{karsenti2001}
{\sc E.~Karsenti and I.~Vernos}, {\em The mitotic spindle: a self-made
  machine}, Science, 294 (2001), pp.~543--547.

\bibitem{kikkawa2001}
{\sc M.~Kikkawa, E.~P. Sablin, Y.~Okada, H.~Yajima, R.~J. Fletterick, and
  N.~Hirokawa}, {\em Switch-based mechanisms of kinesin motors}, Nature, 411
  (2001), p.~439.

\bibitem{platen1992}
{\sc P.~E. Kloeden and E.~Platen}, {\em Numerical solution of stochastic
  differential equations}, Springer-Verlag, 1992.

\bibitem{knight1999}
{\sc A.~E. Knight and J.~E. Molloy}, {\em Coupling atp hydrolysis to mechanical
  work}, Nature Cell Biology, 1 (1999), pp.~E87--E89.

\bibitem{kozielski1997}
{\sc F.~Kozielski, S.~Sack, A.~Marx, M.~Thormahlen, E.~Schonbrum, V.~Biou,
  A.~Thompson, E.~M. Mandelkow, and E.~Mandelkow}, {\em The crystal structure
  of dimeric kinesin and implications for microtubule-dependent motility},
  Cell, 91 (1997), p.~985.

\bibitem{kull1996}
{\sc F.~J. Kull, E.~P. Sablin, R.~Lau, R.~J. Fletterick, and R.~D. Vale}, {\em
  Crystal structure of the kinesin motor domain reveals a structural similarity
  to myosin}, Nature, 380 (1996), pp.~550--555.

\bibitem{landau2000}
{\sc D.~Landau and K.~Binder}, {\em A Guide to Monte-Carlo Simulations in
  Statistical Physics}, Cambridge University Press, 2000.

\bibitem{li1999}
{\sc J.~Li, K.K. Pfister, S.~Brady, and A.~Dahlstrom}, {\em Axonal transport
  and distribution of immunologically distinct kinesin heavy chains in rat
  neurons}, Journal of Neuroscience Research, 58 (1999), pp.~226--241.

\bibitem{maes2003}
{\sc C.~Maes and Maarten~H. van Wieren}, {\em A markov model for kinesin},
  Journal of Statistical Physics, 112 (2003), pp.~329--355.

\bibitem{mandelkow1999}
{\sc E.~Mandelkow and A.~Hoenger}, {\em Structures of kinesin and
  kinesin-microtubule interactions}, Current Opinion in Cell Biology, 11
  (1999), pp.~34--44.

\bibitem{mogilner2001}
{\sc A.~Mogilner, A.J. Fisher, and R.J. Baskin}, {\em Structural changes in the
  neck linker of kinesin explain the load dependence of the motor's mechanical
  cycle}, Journal of Theoretical Biology, 211 (2001), pp.~143--157.

\bibitem{nishiyama2001}
{\sc M.~Nishiyama, E.~Muto, Y.~Inoue, T.~Yanagida, and H.~Higuchi}, {\em
  Substeps within the 8nm step of the atpase cycle of single kinesin
  molecules}, Nature Cell Biology, 3 (2001).

\bibitem{nogales1999}
{\sc E.~Nogales, M.~Whittaker, R.A. Milligan, and K.H. Downing}, {\em
  High-resolution model of the microtubule}, Cell, 96 (1999), pp.~79--88.

\bibitem{oksendal2000}
{\sc B.~Oksendal}, {\em Stochastic Differential Equations: An Introduction with
  Applications}, Springer, 2000.

\bibitem{peskin1995}
{\sc C.~Peskin and G.~Oster}, {\em Coordinated hydrolysis explains the
  mechanical behavior of kinesin}, Biophysics J., 68 (1995), pp.~202--211.

\bibitem{peskin1993}
{\sc C.~S. Peskin, G.~M. Odell, and G.~F. Oster}, {\em Cellular motions and
  thermal fluctuations: The brownian ratchet}, Biophysical Journal, 65 (1993),
  pp.~316--324.

\bibitem{ray1993}
{\sc S.~Ray, E.~Meyh\"{o}fer, R.A. Milligan, and J.~Howard}, {\em Kinesin
  follows the microtubule's protofilament axis}, Journal of Cell Biology, 121
  (1993), pp.~1083--1093.

\bibitem{reichl1998}
{\sc L.~E. Reichl}, {\em A modern course in statistical physics}, John Wiley
  and Sons, 1998.

\bibitem{rice2003}
{\sc S.~Rice, Y.~Cui, S.~Sindelar, N.~Naber, M.~Matuska, R.~Vale, and
  R.~Cooke}, {\em Thermodynamic properties of the kinesin neck-region docking
  to the catalytic core}, Biophysical Journal, 84 (2003), pp.~1844--1854.

\bibitem{rice1999}
{\sc S.~Rice, A.~W. Lin, D.~Safer, C.L. Hart, N.~Naber, B.O. Carragher, S.M.
  Cain, E.. Pechatnikova, E.~W. Wilson-Kubalek, M.~Whittaker, E.~Pate,
  R.~Cooke, E.~W. Taylor, R.A. Milligan, and R.D. Vale}, {\em A structural
  change in the kinesin motor protein that drives motility}, Nature, 402
  (1999), pp.~778--784.

\bibitem{ross1995}
{\sc S.~Ross}, {\em Stochastic Processes}, Wiley Text Books, 1995.

\bibitem{sablin2001}
{\sc E.~P. Sablin and R.~J. Fletterick}, {\em Nucleotide switchets in molecular
  motors: structural analysis of kinesins and myosins}, Current Opinion in
  Structural Biology, 11 (2001), pp.~716--724.

\bibitem{sack1997}
{\sc S.~Sack, J.~Muller, A.~Marx, M.~Thormahlen, E.~M. Mandelkow, S.~T. Brady,
  and E.~Mandelkow}, {\em X-ray structure of motor and neck domains from rat
  brain kinesin}, Biochemistry, 36 (1997), p.~16155.

\bibitem{schliwa2001}
{\sc M.~Schliwa and G.~Woehlke}, {\em Switching on kinesin}, Nature, 411
  (2001), pp.~424--425.

\bibitem{sharp2000}
{\sc D.~Sharp, G.~Rogers, and J.~Scholey}, {\em Microtubule motors in mitosis},
  Nature, 407 (2000), pp.~41--45.

\bibitem{sindelar2002}
{\sc C.V. Sindelar, M.J. Budny, S.~Rice, N.~Naber, R.~Fletterick, and
  R.~Cooke}, {\em Two conformations in the human kinesin power stroke defined
  by x-ray crystallography and epr spectroscopy}, Nature structural Biology, 9
  (2002), pp.~844--848.

\bibitem{song2001}
{\sc Y.~H. Song, A.~Marx, J.~Muller, G.~Woehlke, M.~Schliwa, A.~Krebs,
  A.~Hoenger, and E.~Mandelkow}, {\em Structure of fast kinesin: implications
  for atpase mechanisms and interactions with microtubules}, Embo J., 20
  (2001), p.~6213.

\bibitem{svobodaandblock1994}
{\sc K.~Svoboda and S.~Block}, {\em Force and velocity measured for single
  kinesin molecules}, Cell, 77 (1994), pp.~773--784.

\bibitem{svoboda1994b}
{\sc K.~Svoboda, P.~Mitra, and S.~Block}, {\em Fluctuation analysis of motor
  protein movement and single enzyme kinetics}, Proc. Nat. Acad. Sci.,
  (1994b).

\bibitem{svoboda1993}
{\sc K.~Svoboda, C.~F. Schmidt, B.~J. Schnapp, and S.~M. Block}, {\em Direct
  observation of kinesin stepping by optical trapping interferometry}, Nature,
  365 (1993), pp.~721--727.

\bibitem{tuma1998}
{\sc C.M. Tuma, A.~Zill, N.~Le Bot, I.~Vernos, and V.~Gelfand}, {\em
  Heterotrimeric kinesin ii is the microtubule motor protein responsible for
  pigment dispersion in xenopus melanophores}, The Journal of Cell Biology, 143
  (1998), pp.~1547--1558.

\bibitem{vale1997}
{\sc R.~Vale and R.~Fletterick}, {\em The design plan of kinesin motors}, Annu.
  Rev. Cell Dev. Biol., 13 (1997), pp.~745--777.

\bibitem{visscher1999}
{\sc K.~Visscher, M.~Schnitzer, and S.~M. Block}, {\em Single kinesin molecules
  studies with a molecular force clamp}, Nature, 400 (1999), pp.~184--189.

\bibitem{woehlke1997}
{\sc G.~Woehlke, A.K. Ruby, C.L. Hart, B.~Ly, N.~Hom-Booher, and R.D. Vale},
  {\em Microtubule interaction site of the kinesin motor}, Cell, 90 (1997),
  pp.~207--216.

\bibitem{yun2003}
{\sc M.~Yun, C.~E. Bronner, C.~G. Park, S.~S. Cha, H.~W. Park, and S.~A.
  Endow}, {\em Rotation of the stalk/neck and one head in a new crystal
  structure of the kinesin motor protein}, J. Embo, 22 (2003), p.~1.

\end{thebibliography}

\section*{Appendix}
\subsection*{Potential Energy for the Mechanical Model}

The potential energy for the entire system consists of
\begin{eqnarray}
\mathbf{X}         & = & \left[\mathbf{X}_{bead},\mathbf{X}_{hinge},\mathbf{X}_{h1},
                       \mathbf{X}_{h2} \right] \\
\nonumber
V(\mathbf{X})      & = & V_{tether} + V_{hinge} + V_{motor} + V_{trap} 
\end{eqnarray} 
In this notation the three dimensional positions of the two heads of the 
Kinesin motor are denoted by $\mathbf{X}_{h1}$,$\mathbf{X}_{h2}$.  The bead 
transported by the motor has position
$\mathbf{X}_{bead}$.  The position of the hinge point where the two heads join the 
tether of the motor is given by $\mathbf{X}_{hinge}$.  In the model the energy is 
taken to be a function of only these four degrees of freedom and consists of
contributions from the stretching of the tether, preferred hinge orientation, 
elasticity of the heads, and the optical trap.

The tether potential energy was derived from the experimental data
as explained in section \ref{section_tether_derivation}.  The 
force-extension profile was found to 
be approximated well by a cubic polynomial with coefficients $\{a_k\}$.  
The tether energy is modeled consistently with the cubic fit by the 
following potential.
\begin{eqnarray}
V_{tether}      & = & E_{tether}(|\mathbf{X}_b - \mathbf{X}_{hinge}|) \\
\nonumber
E_{tether}(s)   & = & \frac{a_3}{4} s^4 + \frac{a_2}{3} s^3 
                    + \frac{a_1}{2} s^2 + a_0s 
\end{eqnarray} 

In the model the free head is biased toward the plus end of the microtubule.
This occurs through a force that has the tendency to keep the bound head 
at a fixed angle with the microtubule.  We model this by linear springs
in each spatial direction that tend to maintain the hinge position in a subset
of preferred configurations.  The biasing force is specified by a separate
stiffness for each spatial component.  We remark that in the model 
$K_{bias}^{(3)}$ has been set to zero in all of the simulations but is 
included here for generality.
\begin{eqnarray}
V_{hinge}       & = & 
\frac{K_{bias}^{(1)}}{2}
|\mathbf{X}_{hinge}^{(1)} - (\mathbf{X}_{bound}^{(1)} + \mathbf{x}_0^{(1)})|^2 \\
\nonumber
& + &
\frac{K_{bias}^{(2)}}{2}
|\mathbf{X}_{hinge}^{(2)} - (\mathbf{X}_{bound}^{(2)} + \mathbf{x}_0^{(2)})|^2 \\
\nonumber
& + &
\frac{K_{bias}^{(3)}}{2}
|\mathbf{X}_{hinge}^{(3)} - (\mathbf{X}_{bound}^{(3)} + \mathbf{x}_0^{(3)})|^2 
\end{eqnarray} 

The elasticity of the globular domains of the motor protein that mediate
the force between the microtubule interaction sites of the head and the 
hinge point is modeled by two linear springs that contribute the 
following energy to the potential of the system.
\begin{eqnarray}
V_{motor} & = & \frac{K_m}{2}(|\mathbf{X}_{hinge} - \mathbf{X}_{h1}| - L)^2 
                  + \frac{K_m}{2}(|\mathbf{X}_{hinge} - \mathbf{X}_{h2}| - L)^2 
\end{eqnarray} 

In the simulations the force exerted by the optical trap is modeled by 
a constant load force $\mathbf{F}$ reflecting the small spatial range over 
which we are interested in the experimental setup.  This potential is given by  
\begin{eqnarray}
V_{trap} & = & -\mathbf{F}\cdot{\mathbf{X}_{bead}} 
\end{eqnarray}

\end{document}